\documentclass[letterpaper, prd, aps, superscriptaddress, preprint, tightenlines]{revtex4-1}
\pdfoutput=1
\usepackage{amssymb,amsmath}
\usepackage[english]{babel}
\usepackage{braket}
\usepackage{cleveref}
\usepackage{color}
\usepackage{dsfont}
\usepackage{epsfig}
\usepackage{float}
\usepackage{feynmf}
\usepackage{graphicx}
\usepackage{import}
\usepackage[utf8]{inputenc}
\usepackage{nicefrac}
\usepackage{multirow}
\usepackage{tikz,pgf}
\usepackage{rotating}
\usepackage{siunitx}
\usepackage{subcaption}
\usepackage{textcomp}
\usepackage{todonotes}
\makeatletter
\presetkeys
    {todonotes}
    {inline}{}
\makeatother
\usepackage{wrapfig}

\renewcommand{\l}{\left}
\renewcommand{\r}{\right}

\newcommand{\order}[1]{\mathcal{O}\l({#1}\r)}

\newcommand{\interp}[1]{\mathcal{O}(#1)}
\newcommand{\ma}{\mu_{\pi K}\,a_0}

\newcommand{\chilog}[1]{\ln\frac{#1}{\Lambda_\chi}}
\newcommand{\muazerothreehalves}{\mu_{\pi K}\,a_0^{3/2}}
\newcommand{\muazeroonehalf}{\mu_{\pi K}\,a_0^{1/2}}
\newcommand{\muafinal}{\mu_{\pi K}\,a_0^{3/2,\text{phys}} = -0.0463(17)}
\newcommand{\lpikfinal}{ L_{\pi K} = 0.0038(3)}
\newcommand{\mahigh}{M_{\pi}\,a_0^{3/2,\text{phys}} = -0.059(2)}
\newcommand{\malow}{M_{\pi}\,a_0^{1/2,\text{phys}}=0.163(3)}

\newcommand{\E}[1]{\textbf{E#1}}
\Crefname{appsec}{appendix}{appendices}
\Crefname{appsec}{Appendix}{Appendices}
 
\begin{document}
	\title{\boldmath Hadron-Hadron Interactions from $N_f=2+1+1$ Lattice QCD:\\
  $I=3/2$ $\pi K$ Scattering Length }
	
	\author{C.~Helmes}
	\email[Corresponding author: ]{helmes@hiskp.uni-bonn.de}
        \affiliation{Helmholtz Institut f{\"u}r Strahlen- und Kernphysik,
	  University of Bonn, Bonn, Germany}
	\author{C.~Jost}
        \affiliation{Helmholtz Institut f{\"u}r Strahlen- und Kernphysik,
	  University of Bonn, Bonn, Germany}
	\author{B.~Knippschild}
        \affiliation{Helmholtz Institut f{\"u}r Strahlen- und Kernphysik,
	  University of Bonn, Bonn, Germany}
	\author{B.~Kostrzewa}
        \affiliation{Helmholtz Institut f{\"u}r Strahlen- und Kernphysik,
	  University of Bonn, Bonn, Germany}
	\author{L.~Liu}
        \affiliation{Institute of Modern Physics, Chinese Academy of
          Sciences, Lanzhou, China}
	\author{F.~Pittler}
        \affiliation{Helmholtz Institut f{\"u}r Strahlen- und Kernphysik,
	  University of Bonn, Bonn, Germany}
	\author{C.~Urbach}
        \affiliation{Helmholtz Institut f{\"u}r Strahlen- und Kernphysik,
	  University of Bonn, Bonn, Germany}
	\author{M.~Werner}
	\affiliation{Helmholtz Institut f{\"u}r Strahlen- und Kernphysik,
	University of Bonn, Bonn, Germany}

	\collaboration{\textbf{ETM Collaboration}}
	
	\begin{abstract}
          In this paper we report on results for the s-wave scattering length of the
          $\pi$-$K$ system in the $I=3/2$ channel from $N_f=2+1+1$ Lattice
          QCD. The calculation is based on gauge configurations
          generated by the European Twisted Mass Collaboration with
          pion masses ranging from about
          \SIrange{230}{450}{\mega\electronvolt} at three values of
          the lattice spacing. Our main result reads $\mahigh$. Using
          chiral perturbation theory we are also able to estimate
          $\malow$. The error includes statistical and systematic
          uncertainties, and for the latter in particular errors from
          the chiral and continuum extrapolations.
	\end{abstract}

\maketitle

\section{Introduction}

For understanding the strong interaction sector of the standard model
(SM) it is not sufficient to compute masses of stable
particles. Gaining insight into interactions of two or more
hadrons and resonances is a must. Due to the non-perturbative nature
of low energy quantum chromodynamics (QCD), computations of interaction
properties from lattice 
QCD are highly desirable. While ultimately the phase shift in a given
partial wave is to be computed, also the scattering length is in many
cases a useful quantity, in particular when the two-particle interaction
is weak.

Due to the importance of chiral symmetry in QCD the investigation of
systems with two pseudoscalar mesons is of particular interest. Here,
chiral perturbation theory (ChPT) is able to provide a description of
the pion mass dependence. And any non-perturbative computation in turn
allows to check this dependence. Naturally, ChPT works best for two
pion systems, while convergence is unclear for pion-kaon or two kaon
systems. 

The two pion system is studied well experimentally, also in the
different isospin channels. However, as soon as one or both pions are
replaced by kaons, experimental results become sparse. On the other
hand, this gap starts to be filled by lattice QCD calculations.
For the pion-kaon system with isospin $I=3/2$ there are
by now a few lattice results available focusing on the scattering
length~\cite{Beane:2006gj,Sasaki:2013vxa,Fu:2011wc,Janowski:2014uda,Lang:2012sv}.
The most recent computation in Ref.~\cite{Janowski:2014uda} uses one lattice at
physical pion and kaon masses and lattice spacing
$a\approx\SI{0.114}{\femto\metre}$. For the sea and valence sector they use
$N_f=2+1$ M{\"o}bius domain wall fermions and an Iwasaki gauge action.
In Ref.~\cite{Sasaki:2013vxa} a systematic study of the elastic scattering lengths for
the light pseudoscalar mesons was carried out with $N_f=2+1$ order $\order{a}$-improved
Wilson quarks at pion masses ranging from
\SIrange{170}{710}{\mega\electronvolt} and a lattice spacing
$a\approx\SI{0.09}{\femto\metre}$. Furthermore
Refs.~\cite{Fu:2011wc,Beane:2006gj} use $N_f=2+1$ flavors on the MILC
configurations with a rooted staggered sea quark action. Whereas
Ref.~\cite{Fu:2011wc} calculates the scattering length at a lattice spacing
$a\approx\SI{0.15}{\femto\metre}$, a slightly smaller lattice spacing 
$a\approx\SI{0.125}{\femto\metre}$ has been used in Ref.~\cite{Beane:2006gj}.
The pion masses in Ref.~\cite{Beane:2006gj} range from
\SIrange{290}{600}{\mega\electronvolt} using domain wall valence quarks with a
chiral extrapolation done in mixed-action chiral perturbation theory (MAChPT)~\cite{PhysRevD.73.074510,PhysRevD.75.054501}.
The range of pion masses, \SIrange{330}{466}{\mega\electronvolt}, for the Asqtad
improved staggered fermions of
Ref.~\cite{Fu:2011wc} is a bit smaller compared to Ref.~\cite{Beane:2006gj}.
In Ref.~\cite{Lang:2012sv} the phaseshifts and scattering lengths for
$\pi$-$K$-scattering in $I=3/2$ and $I=1/2$ in the $s$-wave and the $p$-wave has
been determined. The gauge action is a $N_f=2$ tree level improved Wilson-Clover
action. The authors include the strange quark as a valence quark only which then
corresponds to pion and kaon masses of
$M_{\pi}=\SI{266}{\mega\electronvolt}$ and $M_K=\SI{522}{\mega\electronvolt}$,
respectively.

In this paper we are going to present results for the s-wave
scattering length of the pion-kaon system in the elastic region with
isospin $I=3/2$. The investigation is based on gauge configurations
produced by the European Twisted Mass Collaboration (ETMC) with
$N_f=2+1+1$ dynamical quark flavors~\cite{Baron:2010th}. In 
contrast to previous computations, we are able to perform a continuum
extrapolation owing to 11 ensembles with $M_\pi$ ranging from
\SIrange{230}{450}{\mega\electronvolt} distributed over 3 different
lattice spacing values. We employ in total 4 different extrapolation
methods to also estimate systematic uncertainties associated with our
computation.

Finally, since this paper is the fourth in a series of
publications~\cite{Helmes:2015gla,Liu:2016cba,Helmes:2017smr}
concerning elastic scattering of two pions in different channels and
kaon-kaon with $I=1$, we are able to compare results of two
pseudoscalar mesons at maximal isospin involving different amounts of
strangeness. The leading order ChPT predictions for the dependence on
the reduced mass divided by the relevant decay constant are
identical for the three systems and differences appear only at NLO. 

This paper is organized as follows: We first introduce the lattice
details of our calculation. After the discussion of the analysis
methods we present the main result, followed by a detailed discussion
of the analysis details. We close with a discussion and
summary. Technical details can be found in the appendix.
 \section{Lattice Action and Operators}
\subsection{Action}

The lattice details for the investigation presented here are very
similar to the ones we used to study the kaon-kaon scattering
length~\cite{Helmes:2017smr}. We use $N_f=2+1+1$ flavor lattice QCD
ensembles generated by the ETM Collaboration, for which details can be
found in Refs.~\cite{Chiarappa:2006ae,Baron:2010th,Baron:2010bv}. 
The parameters relevant for this paper are compiled in
\Cref{tab:setup}: we give for each ensemble the inverse gauge coupling
$\beta=6/g_0^2$, the bare quark mass parameters $\mu_\ell, \mu_\sigma$
and $\mu_\delta$, the lattice volume and the number of configurations on
which we estimated the relevant quantities.

\begin{table}
 \centering
 \begin{tabular*}{.9\textwidth}{@{\extracolsep{\fill}}lcccccc}
  \hline\hline
  ensemble & $\beta$ & $a\mu_\ell$ & $a\mu_\sigma$ & $a\mu_\delta$ &
  $(L/a)^3\times T/a$ & $N_\mathrm{conf}$  \\ 
  \hline\hline
  A30.32   & $1.90$ & $0.0030$ & $0.150$  & $0.190$  & $32^3\times64$ & $259$  \\
  A40.24   & $1.90$ & $0.0040$ & $0.150$  & $0.190$  & $24^3\times48$ & $376$  \\
  A40.32   & $1.90$ & $0.0040$ & $0.150$  & $0.190$  & $32^3\times64$ & $246$  \\
  A60.24   & $1.90$ & $0.0060$ & $0.150$  & $0.190$  & $24^3\times48$ & $303$  \\
  A80.24   & $1.90$ & $0.0080$ & $0.150$  & $0.190$  & $24^3\times48$ & $300$  \\
  A100.24  & $1.90$ & $0.0100$ & $0.150$  & $0.190$  & $24^3\times48$ & $304$  \\
  \hline                                                                 
  B35.32   & $1.95$ & $0.0035$ & $0.135$  & $0.170$  & $32^3\times64$ & $241$ \\
  B55.32   & $1.95$ & $0.0055$ & $0.135$  & $0.170$  & $32^3\times64$ & $251$ \\
  B85.24   & $1.95$ & $0.0085$ & $0.135$  & $0.170$  & $32^3\times64$ & $288$ \\
  \hline                                                                 
  D30.48 & $2.10$ & $0.0030$ & $0.120$ & $0.1385$ & $48^3\times96$ & $364$ \\
  D45.32sc & $2.10$ & $0.0045$ & $0.0937$ & $0.1077$ & $32^3\times64$ & $289$ \\
  \hline\hline

 \end{tabular*}
 \caption{The gauge ensembles used in this study. For the labeling of
   the ensembles we adopted the notation in
   Ref.~\cite{Baron:2010bv}. In addition to the relevant input
   parameters we give the lattice volume and  the number of evaluated
   configurations, $N_\mathrm{conf}$.}
 \label{tab:setup}
\end{table}

The ensembles were generated using the Iwasaki gauge action and employ the
$N_f=2+1+1$ twisted mass fermion
action~\cite{Frezzotti:2003ni,Frezzotti:2003xj,Frezzotti:2004wz}. 
For orientation, the $\beta$-values \numlist{1.90;1.95;2.10} correspond to
lattice spacing values of $a\sim\,$\SIlist{0.089;0.082;0.062}{\femto\metre},
respectively, see also \Cref{tab:r0values}. The
ensembles were generated at so-called maximal twist, which guarantees
automatic order $\order{a}$ improvement for almost all physical
quantities~\cite{Frezzotti:2003ni}. The
renormalized light quark mass $m_\ell$ is directly proportional to the
light twisted quark mass via
\begin{equation}
  m_\ell\ =\ \frac{1}{Z_P} \mu_\ell\,,
\end{equation}
with $Z_P$ the pseudoscalar renormalization constant. The relation of
the bare parameters $\mu_\sigma$ and $\mu_\delta$ to the renormalized
charm and strange quark masses reads
\begin{equation}
  m_{c,s}\ =\ \frac{1}{Z_P}\mu_\sigma\ \pm\ \frac{1}{Z_S} \mu_\delta\,,
\end{equation}
with $Z_S$ the non-singlet scalar renormalization constant.

As noted in Refs.~\cite{Baron:2010bv, Carrasco:2014cwa}, the renormalized sea strange quark masses across the ``A'', ``B'' and ``D'' ensembles vary by up to about 20\% and in a few cases differ from the physical strange quark mass to the same extent.
For D30.48 and D45.32sc at the finest lattice spacing, the sea strange quark
mass on the former ensemble overshoots the physical strange quark mass while it
is consistent on the latter ensemble.
In order to correct for these mis-tunings and to avoid the complicated
flavor-parity mixing in the unitary non-degenerate strange-charm
sector~\cite{Baron:2010th}, we adopt a mixed action ansatz with so-called
Osterwalder-Seiler (OS)~\cite{Frezzotti:2004wz} valence quarks, while keeping
order $\order{a}$ improvement intact.
We denote the OS bare strange quark parameter with $\mu_s$.
It is related to the renormalized strange quark mass by
\begin{equation}
  m_s\ =\ \frac{1}{Z_P} \mu_s\,.
\end{equation}
For each ensemble we investigate three values of $\mu_s$ which are compiled in \Cref{tab:mus}. 
More details on the mixed action approach can be found in Ref.~\cite{Helmes:2017smr}.

As a smearing and contraction scheme we employ the stochastic Laplacian-Heaviside approach, described in Ref.~\cite{Morningstar:2011ka}.
Details of our parameter choices can be found in Refs.~\cite{Helmes:2015gla,Helmes:2017smr}.

\begin{table}[t!]
 \centering
 \begin{tabular*}{.7\textwidth}{@{\extracolsep{\fill}}lrrr}
  \hline\hline
  $\beta$ & $1.90$ & $1.95$ & $2.10$ \\
  \hline\hline
  $a\mu_s$ & 0.0185 & 0.0160 & 0.013/0.0115\\
           & 0.0225 & 0.0186 & 0.015\\
           & 0.0246 & 0.0210 & 0.018\\
  \hline\hline
 \end{tabular*}
 \caption{Values of the bare strange quark mass $a\mu_s$ used for the
   three $\beta$-values. The lightest strange quark mass on the ensemble D30.48 is $a\mu_s = 0.0115$
   instead of $a\mu_s = 0.013$.}
 \label{tab:mus}
\end{table}

\subsection{Lattice Operators and Correlation Functions}

For reasons which will become clear later we need to estimate the
masses of the pion, the kaon and the $\eta$ meson on our
ensembles. The masses for the pion and kaon are obtained from the large
Euclidean time dependence of two point functions of the form
\begin{equation}
  C_X(t-t') = \langle\interp{X}(t) \, \interp{X}^\dagger(t')\rangle\,,
  \label{eq:two_corr}
\end{equation}
where $X\in\{\pi, K\}$. The operators for the charged pion and kaon
projected to zero momentum read 
\begin{equation}
  \mathcal{O}(X)(t)\ =\ \sum_{\mathbf{x}} O_X(\mathbf{x}, t)
\end{equation}
with
\begin{align}
  O_\pi(\mathbf{x}, t)\ &=\
  i\bar{d}(\mathbf{x},t)\,\gamma_5\,u(\mathbf{x},t)\,,\\
  O_K(\mathbf{x}, t)\ &=\
  i\bar{s}(\mathbf{x},t)\,\gamma_5\,u(\mathbf{x},t)\,.
\end{align}
For the $\eta$ (and $\eta^\prime$) meson we use the two operators
\begin{align}
  O_\ell(\mathbf{x}, t)\ &=\ \frac{i}{\sqrt{2}}(\bar{u}(\mathbf{x},t)\,\gamma_5\, u(\mathbf{x},t) +
  \bar{d}(\mathbf{x},t)\,\gamma_5\, d(\mathbf{x},t))\,,\\
  O_s(\mathbf{x}, t)\ &=\ i\bar{s}(\mathbf{x},t)\,\gamma_5\, s(\mathbf{x},t)\,.
\end{align}
From these we build a two-by-two correlator matrix by taking the
disconnected diagrams into account. The $\eta$
(principal) correlator is determined by solving a generalized
eigenvalue problem as described in detail in
Ref.~\cite{Ottnad:2015hva}.
A complete discussion of the analysis of the $\eta$ (and $\eta'$) meson is beyond the scope of this paper and the full analysis will be presented in a future publication~\cite{Jost:2018nn}.
In addition to the aforementioned meson masses, we also need to
estimate the energy $E_{\pi K}$ of the interacting pion-kaon two particle system.
For the case of maximal isospin, i.e. $I=3/2$, the corresponding
two particle operator reads
\begin{align}
  \interp{\pi K}(t) =
  -\sum_{\mathbf{x},\mathbf{x'}}\bar{d}(\mathbf{x},t)\,\gamma_5\,u(\mathbf{x},t)
  \bar{s}(\mathbf{x'},t)\,\gamma_5\,u(\mathbf{x'},t)\,.
  \label{eq:inter_pik}
\end{align}
It is used to construct the two-particle correlation function
\begin{align}
  C_{\pi K}(t-t') = \langle\interp{\pi K}(t)\interp{\pi K}^{\dagger}(t')\rangle\,.
  \label{eq:four_corr}
\end{align}
$E_{\pi K}$ can then be determined from the large Euclidean time
dependence of $C_{\pi K}$.

 \section{Analysis Methods}
\label{sec:analysis_methods}

We focus in this work on pion-kaon scattering in the elastic region.
For small enough squared scattering momentum $p^2$ one can perform
the effective range expansion for partial wave $\ell$
\begin{equation}
  p^{2\ell+1} \cot(\delta_\ell)\ =\ -\frac{1}{a_\ell} +\mathcal{O}(p^2)
\end{equation}
with phase shift $\delta_\ell$ and scattering length $a_\ell$.
For the pion-kaon system it is, to a very good approximation,
sufficient to study the s-wave, i.e. $\ell=0$.

In lattice QCD the phase shift or the scattering length can only be
computed from finite volume induced energy shifts. The relevant energy
shift here is given by
\begin{equation}
  \delta E = E_{\pi K} - M_\pi - M_K\,.
\end{equation}
Using again the effective range expansion, one arrives at the
L{\"u}scher formula~\cite{Luscher:1986pf} 
\begin{align}
  &\delta E=- \frac{2 \pi a_{0}} {\mu_{\pi K} L^3} \left(1 +c_1 \frac{a_{0}}{L} + c_2
  \frac{a_{0}^2}{L^2}\right) +
  \mathcal{O}(L^{-6})
  \label{eq:scat_lusch}
\end{align}
relating $\delta E$ directly to the scattering length $a_0$,
the reduced mass of the pion-kaon system
\begin{align}
  \mu_{\pi K} = \frac{M_\pi M_K}{M_\pi+M_K}\,,
  \label{eq:red_mass}
\end{align}
and the spatial extent of the finite volume $L$. The 
coefficients read~\cite{Luscher:1986pf}
\[
c_1 =  -2.837297\,,\quad c_2=6.375183\,.
\]
Given $\delta E$, $\mu_{\pi K}$ and $L$, L{\"u}scher's formula allows
one to determine the scattering length $a_0$ by solving \Cref{eq:scat_lusch} for $a_0$.
In what follows, we will describe how we extract $\delta E$ and the
other relevant bare quantities from correlation functions. Then we
will give details on our 
approach to inter- or extrapolate the results to physical conditions
and to the continuum limit.

In order to gain some understanding of systematic uncertainties, we
perform the analysis in two different ways once the bare data has
been extracted.
Combined chiral and continuum extrapolations are performed at fixed strange quark mass using next to leading order ChPT (NLO ChPT) and a variant thereof referred to as the $\Gamma$-method, as described in Ref.~\cite{Beane:2006gj}.

\subsection{Physical Inputs}

\begin{table}
  \begin{tabular*}{.5\linewidth}{@{\extracolsep{\fill}}lrr}
    \hline\hline
    $\beta$ &
    $a\ [\mathrm{fm}]$ & $r_0/a$ \\
    \hline\hline
    $1.90$  & $0.0885(36)$ & $5.31(8)$ \\
    $1.95$  & $0.0815(30)$ & $5.77(6)$ \\
    $2.10$  & $0.0619(18)$ & $7.60(8)$ \\
    \hline\hline
  \end{tabular*}
  \caption{Values of the Sommer parameter $r_0/a$ and the
    lattice spacing $a$ 
    at the three values of $\beta$. See 
    Ref.~\cite{Carrasco:2014cwa} for details.}
  \label{tab:r0values}
\end{table}

For the analysis presented below, we require physical inputs for the
pion, the kaon and $\eta$-meson masses as well as the pion decay constant.
To this end, we employ the values in the isospin symmetric limit, $\overline{M}_{\pi}$ and $\overline{M}_{K}$, as determined in chiral perturbation theory~\cite{Gasser:1983yg} and given in Ref.~\cite{Aoki:2016frl} as
\begin{equation}
\begin{split}
  \overline{M}_{\pi} & = 134.8(3)\,\text{MeV}\,, \\
  \overline{M}_{K} & = 494.2(3)\,\text{MeV} \,.
\end{split}
\label{eq:meson_masses_phys}
\end{equation}
For the $\eta$ meson mass we use the average obtained by the Particle Data
Group~\cite{PhysRevD.98.030001}:
\begin{equation}
  \label{eq:etamass}
  \overline{M}_\eta = 547.86(2) \,\text{MeV}\,.
\end{equation}
For the decay constant, we use the phenomenological average determined by the Particle Data Group given in Ref.~\cite{Olive:2016xmw} as
\begin{equation}
  f_{\pi^{-}}^{\text{(PDG)}}  = 130.50(13)\,\text{MeV}\,.
  \label{eq:decay_const_phys}
\end{equation}

As an intermediate lattice scale, we employ the Sommer parameter
$r_0$~\cite{Sommer:1993ce}. It was determined in
Ref.~\cite{Carrasco:2014cwa} from the ensembles we use here to be
\begin{align}
  r_0 = \SI{0.474+-0.011}{\femto\metre}\,.
  \label{eq:r0_cont}
\end{align}
In the parts of the analysis which require $r_0$, we use
parametric bootstrap samples with central value
and width given in \Cref{eq:r0_cont}.
Where $r_0/a$ values enter as  fit parameters, we constrain the corresponding
fit parameters using Gaussian priors in the augmented $\chi^2$ function given as
\begin{equation}
  \chi^2_\mathrm{aug}\ =\ \chi^2 +  \sum_\beta
  \left(\frac{(r_0/a)(\beta) - P_{r}(\beta)}{\Delta
      r_0/a(\beta)}\right)^2\,.
	\label{eq:chi_aug}
\end{equation}

\subsection{Energy Values from Correlation Functions}
\label{sec:pollution}

The energies of the two point correlation functions as given in \Cref{eq:two_corr} are
extracted from fits of the form
\begin{equation}
  C_X(t) = A_0^2(e^{-E_{X}t}+e^{-E_X(T-t)})\,,
  \label{eq:c2_fit}
\end{equation}
to the data.
While for $M_K$ and $M_\pi$ the signal extends up to $T/2$, for the $\eta$ we have to face more noise .
We deal with this by applying the excited state subtraction method used and described in Refs.~\cite{Michael:2013gka,Ottnad:2015hva}.

In the determination of the energy shift $\delta E$, the total energy $E_{\pi K}$
of the interacting $\pi$-$K$ system must be computed.
However, in the spectral decomposition of the two-particle correlation function,
unwanted time dependent contributions, so-called thermal pollutions, appear.
Taking into account that our $\pi$-$K$ correlation function is symmetric around the $T/2$ point, the leading contributions in the spectral decomposition can be cast into the form
\begin{equation} \label{eq:pollution}
\begin{split}
  C_{\pi K}(t) &\approx \braket{\Omega|\pi^+K^+|\pi K}\braket{\pi K|(\pi^+ K^+)^{\dagger}|\Omega}\left(e^{-E_{\pi K}t}+e^{-E_{\pi K}(T-t)}\right) \\
        &+ A_1 \left(e^{-E_{\pi} T}e^{(E_{\pi}-E_K)t} + e^{-E_K T}e^{(E_{K}-E_{\pi})t} \right) \,, 
\end{split}
\end{equation}
where only the first line corresponds to the energy level we are interested in.
However, at finite $T$-values, the second contribution might be
sizable, in particular at times close to $T/2$.
Moreover, the thermal pollution cannot be separated easily from the signal we are interested in.
We have studied two different methods, labeled \E{1} and \E{2}, to extract $E_{\pi K}$ from $C_{\pi K}(t)$, where \E{1} has already been discussed in Ref.~\cite{Dudek:2012gj}.
\begin{itemize}
\item {\E{1}: weighting and shifting:}\\
To render one of the polluting terms
in \Cref{eq:pollution} time independent, the correlation function first gets
weighted by a factor $\exp((E_K - E_\pi)t)$. We chose this factor,
because $\exp(-E_\pi T)$ is significantly larger than $\exp(-E_K T)$.
The resulting constant term can then be removed by the shifting procedure,
which thus replaces $C_{\pi K}(t)$ by
\begin{equation}
\begin{split}
  C^w_{\pi K}(t) &= e^{(E_K-E_\pi)t} C_{\pi K}(t)\,, \\
  \widetilde{C}^w_{\pi K}(t) &= C^w_{\pi K}(t) - C^w_{\pi K}(t+\delta t) \,,
\end{split}
\end{equation}
where $\delta t$ is a fixed number of time slices.

Subsequently, we multiply $\widetilde{C}^w_{\pi K}(t)$ by
$\exp(-(E_K-E_\pi)t)$, which (mostly) recovers the original time
dependence in the contribution of interest
\begin{align}
  C^{\text{E1}}_{\pi K}(t) =e^{-(E_K - E_\pi)t}\widetilde{C}^w_{\pi K}(t)\,.
  \label{eq:reweight}
\end{align}
We then extract the total energy of the $\pi$-$K$ system, $E_{\pi K}$, from a fit of the form,
\begin{equation}
\begin{split}
  C^{\text{E1}}_{\pi K}(t) &= A_0^2\left(e^{-E_{\pi K}t}+e^{-E_{\pi
      K}(T-t)}-e^{(E_K - E_\pi)
	\delta t}\left(e^{-E_{\pi K}(t+\delta t)}+e^{-E_{\pi K}(T-(t+\delta
t))}\right)\right) \\
	&+\widetilde{A}_1 e^{(E_{K}-E_{\pi})t}\,.
	\label{eq:e_fit}
\end{split}
\end{equation}
Note that in contrast to Ref.~\cite{Dudek:2012gj}, where correlator matrices with various thermal pollutions are considered, we are able to take $\tilde{A}_1$ as an additional fit parameter in order to account for this sub-leading term.

\item {\E{2}: dividing out the pollution:}\\
To improve on method \E{1}, we assume that the decomposition given in \Cref{eq:pollution} allows one to neglect any further thermal pollutions.
This leads to dividing out the time dependent part
\begin{align}
  p(t) = e^{(E_K-E_\pi)t}e^{-E_KT}+e^{-(E_K-E_\pi)t}e^{-E_\pi T}\,,
  \label{eq:c4_poll}
\end{align}
explicitly. With
\begin{equation}
  C'_{\pi K}(t)\ =\ \frac{C_{\pi K}(t)}{p(t)}
\end{equation}
we then proceed to calculate
\begin{align}
  &\widetilde{C}_{\pi K}(t) = C'_{\pi K}(t)-C'_{\pi K}(t+\delta t)\,,\\
  &C^{\text{E2}}_{\pi K}(t) = p(t)\widetilde{C}_{\pi K}(t)\,,
  \label{eq:e2_meth}
\end{align}
from which $E_{\pi K}$ can be extracted through a fit of the form
\begin{align}
C^{\text{E2}}_{\pi K}\left(t\right)=A_{0}^2\left(e^{-E_{\pi K}t}+e^{-E_{\pi K}\left(T-t\right)}-\frac{p\left(t\right)}{p\left(t+1\right)}\cdot
\left(e^{-E_{\pi K}\left(t+1\right)}+e^{-E_{\pi K}\left(T-\left(t+1\right)\right)}\right)\right) \,.
\label{eq:e2_fit}
\end{align}
\end{itemize}
We remark that for both methods \E{1} and \E{2} the energies $E_\pi$
and $E_K$, i.e. $M_\pi$ and $M_K$ for zero momentum, are required as an input.
They are determined from the corresponding two-point correlation functions.
For the error analysis bootstrap samples are used to fully preserve all correlations.

\label{sec:method_extra}

After solving \Cref{eq:scat_lusch} for $a_0$ up to order $\order{L^{-5}}$ on every ensemble for each strange quark mass of \Cref{tab:mus}, we have three parameters in which we want to extra- or interpolate: the lattice spacing $a$, the strange quark mass $m_s$ and the light quark mass $m_\ell$.
To evaluate $a_0$ at the physical point we follow a two step procedure.
We first fix the strange quark mass to its physical value and do a combined chiral and continuum extrapolation afterwards.
For the extrapolation in the light quark mass, we use the continuum ChPT expression in terms of meson masses and decay constants, as detailed in Refs.~\cite{Kubis:2001ij,PhysRevD.75.054501,Fu:2011wc}. 

\subsection{Fixing the strange quark mass}
\label{sec:strange_quark}

In order to fix the strange quark mass we adopt the following procedure:
we match the quantity
\begin{align}
  M_s^2=M_K^2-0.5 M_\pi^2\,,
  \label{eq:fix_A}
\end{align}
which is proportional to the strange quark mass at the leading order of ChPT, to its physical value
\begin{equation}
  (M_s^{\text{phys}})^2\ =\ \overline{M}_K^2 - 0.5 \overline{M}_{\pi}^2 \,,
 \label{eq:fix_A_phys}
\end{equation}
using our determinations of $M_K^2$ at three valence strange quark masses on a per-ensemble basis.
For each ensemble, we then interpolate all valence strange quark mass dependent
observables, i.e. $\ma^{3/2}$, $M_K$, $M_{\eta}$ and $\mu_{\pi K}$, in $M_s^2$ to this reference value.

\subsection{Chiral extrapolation}

With the strange quark mass fixed, the extrapolation to the physical point can
be carried out using ChPT. The first NLO calculation of the scattering
amplitude and scattering lengths was done in Ref.~\cite{Bernard:1990kw}.  
From the formulae for the isospin even (odd) scattering lengths $a^+$
($a^-$) in Ref.~\cite{Kubis:2001ij} the NLO ChPT
formulae for $\ma^{I}\,,I\in\Set{1/2,3/2}$, can be derived as sketched in \Cref{app:pik_chpt}, giving
\begin{equation}
  \begin{split}
    \muazerothreehalves &= \frac{\mu^2_{\pi K}}{4\pi f_\pi^2}
    \left[\frac{32M_\pi M_K}{f_\pi^2} L_{\pi K}(\Lambda_\chi)-1-
      \frac{16 M_\pi^2}{f_\pi^2}L_{5}(\Lambda_\chi)\right. \\
      &\left.+\frac{1}{16\pi^2 f_\pi^2} \chi_{\text{NLO}}^{3/2}(\Lambda_\chi,M_\pi,M_K,M_\eta)\right] +c\cdot f(a^2)\,.
  \end{split}
  \label{eq:a0_alg_fit}
\end{equation}
\Cref{eq:a0_alg_fit} depends on the masses of the pion and the kaon, their reduced mass as defined in \Cref{eq:red_mass}, the $\eta$ mass and the pion decay constant.
In addition, the equation depends on the low energy constants (LECs) $L_5$ and $L_{\pi K}$ while $\chi_{\text{NLO}}^{3/2}$ is a known function, see \Cref{app:nlo}.

We express \Cref{eq:a0_alg_fit} in terms of the meson masses and decay constants as they are determined on the lattice, which has the benefit that their ratios can be determined with high statistical precision without the need for explicit factors of the lattice scale.
Formally we fix the scale-dependent LECs at the renormalization scale $\Lambda_\chi = f_{\pi^-}^\text{(PDG)}$.
However, in practice we employ $a \Lambda_\chi = a f_\pi(\beta,\mu_\ell) / K^\text{FSE}_{f_\pi}$  in all chiral logarithms, where the values for the finite-size correction factor $K^\text{FSE}_{f_\pi}$ are given in \Cref{sec:meson_masses_eshift_scat_length}.
Doing so should only induce higher order corrections in the chiral expansion.

Automatic order $\order{a}$ improvement of Wilson twisted mass fermions at
maximal twist guarantees that the leading lattice artifacts are of
order $\order{a^2}$ or better.
For instance, for the $I=2$ $\pi \pi$ s-wave scattering length, discretization
effects start only at order $\order{a^2 M_\pi^2}$~\cite{Buchoff:2008hh}.
A corresponding theoretical result for $\pi K$ is missing so far.
However, our numerical data suggest that also for $\pi K$ lattice artefacts are very small.
Still, we include a term $c\cdot f(a^2)$ accounting for possible
discretization effects, with fit parameter $c$ and $f(a^2)$ either equal to
$a^2/r_0^2$ or to $a^2 M_X^2$, with $M_X^2$ one of the masses or mass
combinations $M_{\pi}^2\,,\,M_K^2, M_K^2+0.5M_{\pi}^2, \mu_{\pi K}^2$. In the
following analysis we will include the term $c\cdot f(a^2)$ into our fit for
every choice of $f(a^2)$ and thus investigate a possible dependence of our data
on the lattice spacing.

To summarize, our fit parameters are the LECs $L_5$ and $L_{\pi K}$, and $c$, where $L_{\pi K}$ is the combination of renormalized LECs
\begin{align}
  L_{\pi K} = 2L_1+2L_2+L_3-2L_4-\frac{L_5}{2}+2L_6+L_8\,.
  \label{eq:lec_lin_comb}
\end{align}
Let us mention already here that the fits to the data described in the next section turn out to be not sensitive to $L_5$.
Therefore, we include it as a prior in the fit with the value taken from Ref.~\cite{Aoki:2016frl}.
In slight abuse of language we will denote this extrapolation method
as NLO ChPT.

\subsection{Extrapolations Using the $\Gamma$ Method at Fixed $m_s$}
\label{sec:lin_chpt}

Next, we describe an alternative way to extrapolate our data.
For this we fix the strange quark mass as described in the previous subsection and interpolate the other quantities.
Using the interpolated data we can compute the following quantity~\cite{Beane:2006gj}
\begin{align}
  \Gamma\left(\frac{M_\pi}{f_\pi},\,\frac{M_K}{f_\pi}\right) =
  -\frac{f^2_\pi}{16M^2_\pi}\left(\frac{4\pi f^2_\pi}{\mu^2_{\pi
      K}}[\ma^{3/2}] + 1
  +\chi_{\text{NLO}}^{-}-2\frac{M_KM_\pi}{f^2_\pi}\chi_{\text{NLO}}^{+}\right)\,,
  \label{eq:gamma_calc}
\end{align}
with next to leading order ChPT functions $\chi^{\pm}_{\text{NLO}}$
given in \Cref{app:nlo}.
Using the ChPT expressions for the isospin even/odd scattering lengths $a^+$ and $a^-$, $\Gamma$ can also be expressed as~\cite{Beane:2006gj}
\begin{align}
  \Gamma\left(\frac{M_\pi}{f_\pi},\,\frac{M_K}{f_\pi}\right) =
  L_5-2\frac{M_K}{M_\pi}L_{\pi K}\,.
  \label{eq:gamma_fit}
\end{align}
For the derivation see \Cref{app:pik_chpt}.
Hence, a linear fit of \Cref{eq:gamma_fit} to the data obtained via \Cref{eq:gamma_calc} allows one to determine the LECs $L_5$ and $L_{\pi K}$.
Given $L_5$ and $L_{\pi K}$ from the fit one can compute $\ma^{3/2}$ at the physical point using \Cref{eq:a0_alg_fit}.
Again, it turns out we are not sensitive to $L_5$ in our fits.
Therefore, we use a prior as discussed before.
This extrapolation method we denote as $\Gamma$ method.

 \section{Results}
\label{sec:results}

\begin{figure}[t]
  \includegraphics{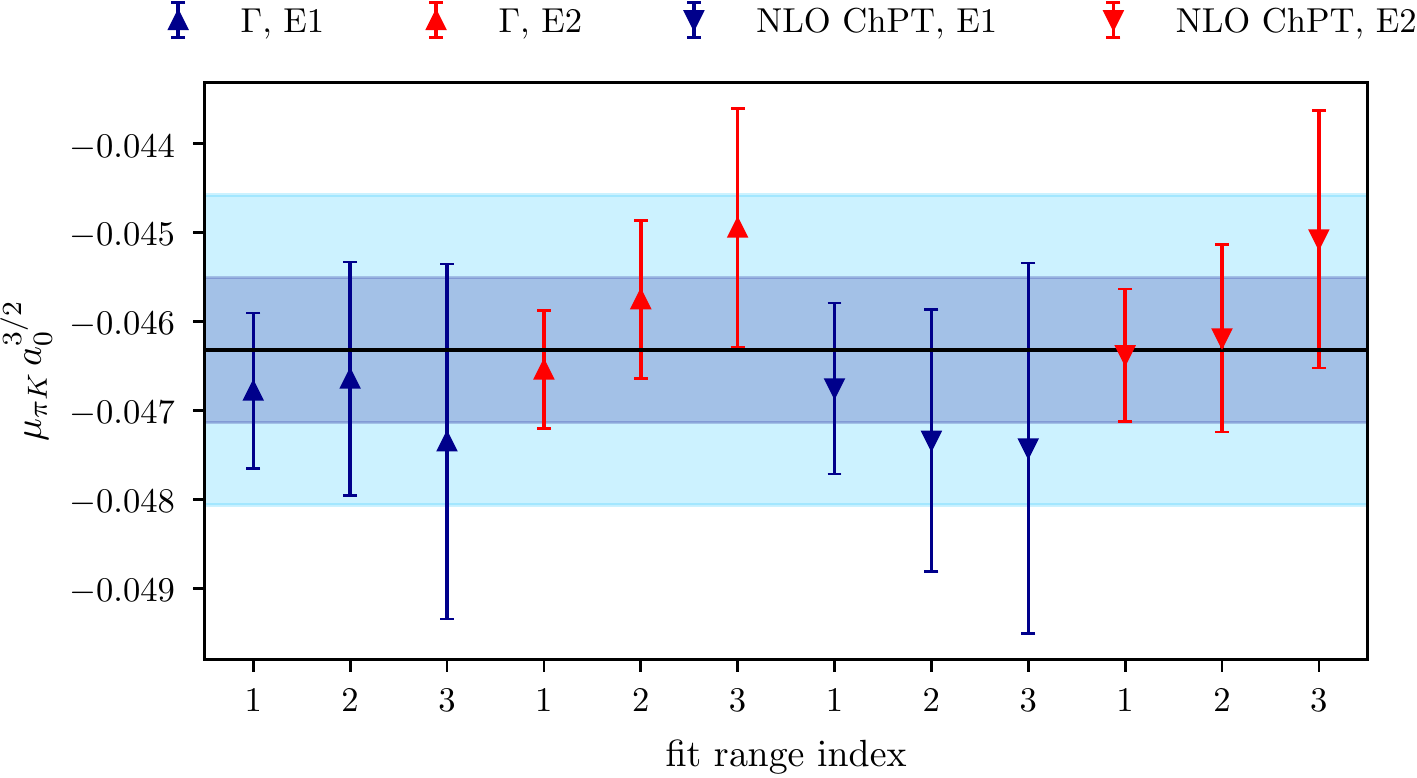}
  \caption{Comparison of values for $\ma^{3/2}$ in the continuum
    limit at the physical point obtained with the different methods
    used in this paper. The fit ranges decrease with increasing index as
    described in \Cref{tab:chpt_ranges}. The inner error band
    represents the statistical error only, while the outer error band
    represents the statistical and systematic errors added in quadrature.}
  \label{fig:methods}
\end{figure}

In this section we present our main result for $\ma^{3/2}$ in the continuum limit and extrapolated to the physical point.

We use two thermal state pollution removal methods, \E{1} and \E{2}, for
$E_{\pi K}$. Next we employ the two (related) ChPT extrapolations,
$\Gamma$-method and NLO ChPT, as discussed before. For each of the two
ChPT extrapolation methods we use three fit ranges as compiled in
\Cref{tab:chpt_ranges}. Hence, we have twelve estimates for each quantity
at the physical point in the continuum limit available, which we use
to estimate systematic uncertainties. We remark that the fit for the
$\Gamma$ method is in terms of $M_K/M_\pi$ and for NLO ChPT in terms
of $\mu_{\pi K}/f_\pi$. Thus, we vary the fit range at the lower end
for the $\Gamma$ method and at the upper end for NLO ChPT.

For $\ma^{3/2}$ the twelve estimates are shown in
\Cref{fig:methods}. The final result is obtained as the weighted average over all of these,
as shown in the figure as the horizontal bold line. The
weight is computed according to
\begin{equation}
  w = \frac{(1- 2 \cdot |p - 1/2|)^2}{\Delta^2}  
\end{equation}
with $p$ the $p$-value of the corresponding ChPT fit and $\Delta$ the
statistical uncertainty obtained from the fit.

The statistical uncertainty of the final results is determined from
the bootstrap procedure. For $\ma^{3/2}$ this is shown in
\Cref{fig:methods} as the inner error band.
In addition, we determine three systematic uncertainties: The first is
obtained from the difference between using only \E{1} or only \E{2}
results. The second from the difference between using only the $\Gamma$
method or only NLO ChPT. Finally, we use the maximal difference of the
weighted average to the twelve estimates as a systematic uncertainty
coming from the choice of fit ranges.

The results of all twelve fits can be found in \Cref{tab:lin_chpt_res_ex}
for the $\Gamma$ method and \Cref{tab:nlo_chpt_res_ex} for NLO ChPT fits.
The fit range indices used in \Cref{fig:methods} are resolved in
\Cref{tab:chpt_ranges}.

\begin{table}
  \centering
  \begin{tabular*}{.5\textwidth}{@{\extracolsep{\fill}}lrrr}
    \hline\hline
    Method & index & Begin & End \\
    \hline
    \multirow{3}{*}{$\Gamma$} &1 &1.2 &2.0 \\
    &2 &1.4 &2.0 \\
    &3 &1.5 &2.0 \\
    \hline
    \multirow{3}{*}{NLO} &1 &1.2 &1.6 \\
    &2 &1.2 &1.41 \\
    &3 &1.2 &1.35 \\
    \hline\hline
  \end{tabular*}
  \caption{Fit ranges used for extrapolations $\Gamma$ and NLO ChPT. The index
    column refers to \Cref{fig:methods}.}
  \label{tab:chpt_ranges}
\end{table}

With this procedure and all
errors added in quadrature we quote
\begin{align}
  \muafinal\,,\qquad
  \lpikfinal\,.
\end{align}
This translates to
\begin{align}
  \mahigh\,,\qquad
  \malow
\end{align}
as our final results. The error budget is compiled in
\Cref{tab:errorbudget}.
While the dominating contribution to the error for both $\ma^{3/2}$
and $L_{\pi K}$
summed in quadrature is coming from the fit range and the statistical
uncertainty, also the choice of the thermal state removal method
contributes significantly.
The contribution from the different chiral extrapolation methods is negligible.
If the errors were added (not in quadrature),
the total error would become a factor $\sim1.7$ larger.

We remark that these results have been obtained with $L_5$ as an
input, because the fits are not sufficiently sensitive to determine
$L_5$ directly. We use the most recent determination from a
$N_f=2+1+1$ lattice calculation by HPQCD~\cite{Aoki:2016frl}, which is
extrapolated to the continuum limit. At our renormalization scale it reads
\begin{equation}
  L_5\ =\ \num{5.4(3)e-3}\,.
  \label{eq:l5_prior}
\end{equation}

\begin{table}[ht]
  \centering
  \begin{tabular*}{.7\textwidth}{@{\extracolsep{\fill}}lrr}
    \hline\hline
    & $\ma^{3/2}\cdot 10^5$ & $L_{\pi K}\cdot 10^5$ \\
    \hline
    statistical                 & $82\ (28\%)$	& $15\ (32\%)$\\
    fit range										& $139\ (47\%)$	&	$19\ (41\%)$\\
    \E{1} v. \E{2}              & $64\ (22\%)$  & $12\ (24\%)$\\
    NLO ChPT v. $\Gamma$        & $9\ (3\%)$  & $1\ (3\%)$\\
    \hline
    $\sum$                      & $294\ (100\%)$ & $47\ (100\%)$ \\
    sqrt $\sum$ in quadrature   &  173            &27  \\
    \hline\hline
  \end{tabular*}
  \caption{Error budget for the final results of $\ma^{3/2}$ and
    $L_{\pi K}$. }
  \label{tab:errorbudget}
\end{table}

 \section{Analysis Details and Discussion}

\subsection{Error Analysis, Thermal Pollution and Choice of Fit Ranges}

The error analysis is performed using the stationary blocked bootstrap
procedure~\cite{Dimitris:1994}.
In order to determine an appropriate average block length, we compute the integrated autocorrelation time $\tau_{\mathrm{int}}$ for the correlation functions $C_{X}(t)$ at all source-sink separations, with $X$ being $\pi$, $K$, $\eta$ or $\pi K$.
In the case of $\pi K$, $C_{X}(t)$ is of course first suitably transformed for the extraction of the interaction energy as discussed in \cref{sec:pollution}.
The computation of $\tau_{\mathrm{int}}$ is detailed in Ref.~\cite{Wolff:2003sm}.
The average block length is then chosen to be the ceiling of the maximum integrated autocorrelation time
observed over all correlation functions at all source-sink separations
\begin{displaymath}
  b=\lceil \max_{X,t} \left( \tau^{(X,t)}_{\mathrm{int}} \right) \rceil
\end{displaymath}
on a per-ensemble basis.
We have confirmed explicitly that this method produces a block length at which the
estimate of the statistical error
plateaus and are thus confident that we properly take into account
the effect of autocorrelations on our quoted statistical errors.
Using the so-determined block length on a per-ensemble basis, 
we generate $N=1500$ samples from which
we estimate statistical errors throughout our analysis.

As discussed in \Cref{sec:pollution}, we employ methods \E{1} and \E{2} to remove unwanted thermal pollutions from the $\pi K$ two particle correlation function.
Both methods allow us to describe the data rather well, but the choice of best fit range depends on the method used to remove the thermal pollution.
This in turn affects the value of the extracted $E_{\pi K}$ and, subsequently, the value of $\muazerothreehalves$ obtained from the energy difference.

\begin{figure}
  \begin{subfigure}{0.49\textwidth}
    \includegraphics{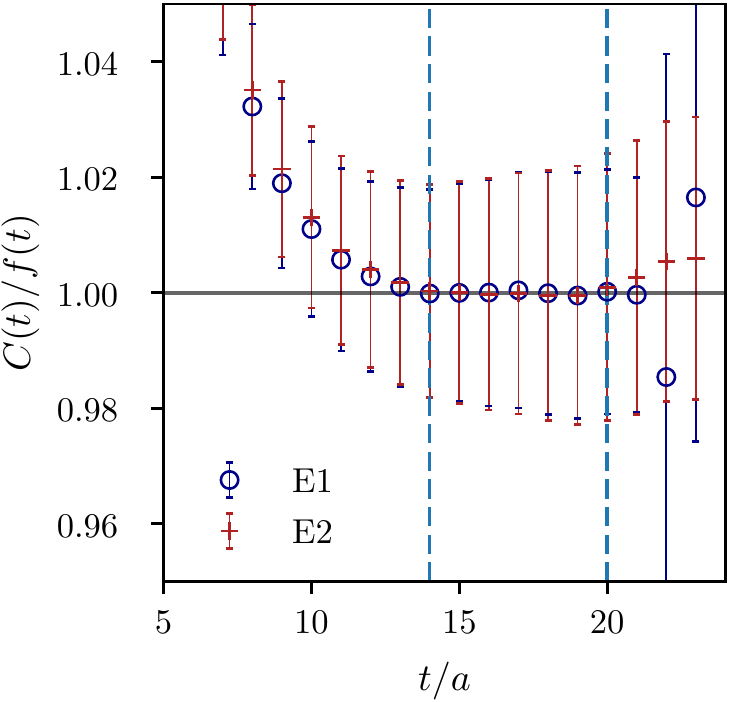}
    \caption{Fit range: $[14,20]$ 
      \\ $aE_{\pi K} = \lbrace 0.389(1)^{\text{E1}}, 0.389(1)^{\text{E2}} \rbrace$} 
  \end{subfigure}
  \begin{subfigure}{0.49\textwidth}
    \includegraphics{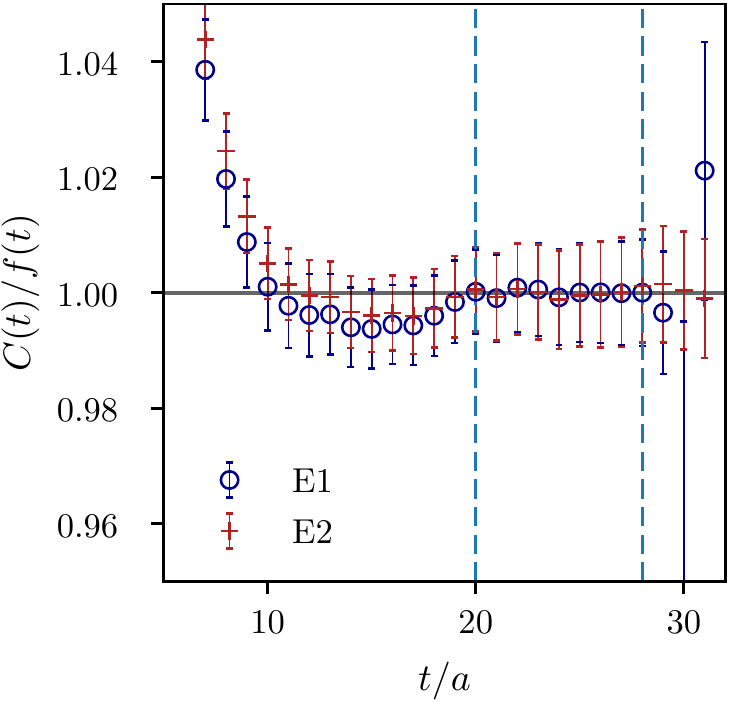}
    \caption{Fit range: $[20,28]$
      \\$aE_{\pi K} = \lbrace 0.3793(6)^{\text{E1}}, 0.3789(5)^{\text{E2}} \rbrace$}
  \end{subfigure}
  \caption{Plot of \Cref{eq:C_over_fit_compare} for ensembles A40.24 and A40.32
    for the lightest strange quark mass for the fit ranges used for the analysis, comparing the quality of the data description by methods \E{1} and \E{2}.}
  \label{fig:C_over_fit_compare}
\end{figure}

To demonstrate the quality of our fits, we look at the ratio
\begin{equation}
  \frac{C^{\text{[E1,E2]}}_{\pi K}(t)}{f^{\text{[E1,E2]}}(t) } \,,
  \label{eq:C_over_fit_compare}
\end{equation}
where $C^{\text{[E1,E2]}}$ are defined in
\Cref{eq:reweight,eq:e2_meth}, respectively, and the fit functions
$f^{\text{[E1,E2]}}(t)$ are given in \Cref{eq:e_fit,eq:e2_fit}, respectively.
The ratio is shown in \Cref{fig:C_over_fit_compare} for the two ensembles A40.24 and A40.32 for two fit ranges for which both methods describe the data well.

The choice of the fit ranges to determine energy levels is always
difficult. In the past, we have used many fit ranges and weighted them
according to their fit
qualities~\cite{Helmes:2015gla,Helmes:2017smr}. However, this
procedure relies on properly estimated variance-covariance matrices,
which is notoriously difficult. For the pion-kaon correlation
functions needed in this paper we have observed several cases where
the fit including the variance-covariance matrix did not properly
describe the data after visual inspection. Therefore, we use fits here
assuming independent data points with the correlation still taken into
account by the bootstrap procedure.

As a consequence, we cannot apply the weighting procedure used in
Refs.~\cite{Helmes:2015gla,Helmes:2017smr} any longer
and have to choose fit ranges. The procedure is as follows:
Due to exponential error growth of $C_{\pi K}$ we fix $t_f=T/2-4a$ and vary
$t_i$, beginning from a region where excited states do not contribute
significantly anymore. From these fits we choose one fit range where the ratio of
\Cref{eq:C_over_fit_compare} is best compatible with 1. The statistical error is
calculated from the bootstrap samples as discussed before. We then estimate the systematic
uncertainty from the remaining fit ranges. To this end we determine the
difference of the mean value to the upper and lower bound of values for $E_{\pi
  K}$. This procedure results in an asymmetric estimate of the systematic
uncertainty of $E_{\pi K}$. The results are compiled in \Cref{tab:raw_data_energies_comp}.
Since $C_K$ and $C_\pi$ do not suffer from exponential error growth at late times we set $t_f=T/2$.
\Cref{tab:fit_ranges} gives an overview of our chosen values of $t_i$ and $t_f$ for all ensembles.
\begin{table}
  \centering
  \begin{tabular*}{.6\textwidth}{@{\extracolsep{\fill}}lllllll}
    \hline\hline
 &      \multicolumn{2}{c}{$aM_\pi$} & \multicolumn{2}{c}{$aM_K$} & \multicolumn{2}{c}{$aE_{\pi K}$} \\
\hline
Ensemble        &       $t_i$ & $t_f$ & $t_i$ & $t_f$ & $t_i$ & $t_f$\\
\hline
A30.32  & 13    & 32    & 13    & 32    & 21    & 28\\ 
A40.24  & 11    & 24    & 11    & 24    & 14    & 20\\
A40.32  & 13    & 32    & 13    & 32    & 20    & 28\\
A60.24  & 11    & 24    & 11    & 24    & 16    & 20\\
A80.24  & 11    & 24    & 11    & 24    & 16    & 20\\
A100.24 & 11    & 24    & 11    & 24    & 15    & 20\\
B35.32  & 13    & 32    & 13    & 32    & 22    & 28\\
B55.32  & 13    & 32    & 13    & 32    & 19    & 28\\
B85.24  & 11    & 24    & 11    & 24    & 15    & 20\\
D45.32sc  & 14    & 32    & 14    & 32    & 22    & 28\\
D30.48  & 22    & 48    & 22    & 48    & 35    & 44\\
\hline\hline
  \end{tabular*}
  \caption{Typical minimal and maximal values of the starting and end points of fit
  ranges for the Correlation functions under investigation.} 
  \label{tab:fit_ranges}
\end{table}

It is always difficult to include systematic uncertainties in the
analysis chain. Since we see systematic uncertainties on extracted
energies on the same level as the statistical one, we adopt the
following procedure to include this uncertainty:
Because $\ma^{3/2}$ is derived from $E_{\pi K}$ we chose to scale
the statistical error for $\ma^{3/2}$ on each ensemble after the data have been
interpolated in the strange quark mass. To this end we define a scaling
factor $s$ via the standard error $\Delta X$ and the average of the
systematic uncertainties $\overline{Q}_X$ over the 3 strange quark
masses for each ensemble:
\begin{align}
  s =
  \sqrt{ \frac{ (\Delta X)^2+\overline{Q}_X^2}{(\Delta X)^2} }\,,
  \label{eq:scalefac_err}
\end{align}
where the average $\overline{Q}_X$ is the simple mean over the six
systematic errors.

\subsection{\boldmath Meson Masses, Energy Shift $\delta E$ and
  Scattering Length $\ma^{3/2}$}
\label{sec:meson_masses_eshift_scat_length}

In order to extract $\delta E$ we first determine $M_K$ and $M_\pi$ from fitting \Cref{eq:c2_fit} to our data for $C_{\pi}(t)$ and $C_{K}(t)$.
We then calculate the reduced mass $\mu_{\pi K}$ via \Cref{eq:red_mass} for all combinations of fit ranges.
$M_\pi$, $M_K$ and $\mu_{\pi K}$ are listed in \Cref{tab:raw_data_masses_comp}. 

The two methods \E{1} and \E{2} give us two estimates of $E_{\pi K}$ as outlined in \Cref{sec:pollution}, from which we determine $\delta E$ and hence the scattering length using \Cref{eq:scat_lusch}.
The values for $E_{\pi K}$ and $\delta E$ are collected in \Cref{tab:raw_data_energies_comp,tab:raw_data_deltae_comp}.

We introduce factors $K^{\text{FSE}}_X$ for
$X~\in~\{M_K,\,M_\pi,\,f_\pi\}$ to correct our lattice data for finite size
effects. They have already been calculated in
Ref.~\cite{Carrasco:2014cwa} and are listed in \Cref{tab:ext_dat}. We apply
these factors for e.g. $M_\pi$ like
\[
M^*_\pi=\frac{M_\pi}{K^{\text{FSE}}_{M_\pi}}\,.
\]
We correct every quantity of the set named above and drop the asterisk in what follows to improve legibility.
For $M_\eta$ statistical uncertainties are too big to resolve finite
volume effects, see also Ref.~\cite{Ottnad:2017bjt}.

For the two methods \E{1} and \E{2} we solve \Cref{eq:scat_lusch} for $a_0$ up to order
$\mathcal{O}(L^{-5})$ numerically.
The values for $a_0$ and its product with the reduced mass, $\ma^{3/2}$ are collected
in \Cref{tab:raw_data_scatlen_comp}.
Since the finite size behavior of the scattering length is unknown, we do not
apply finite size corrections to the reduced mass appearing in $\ma^{3/2}$, either. 

\subsection{Strange quark mass fixing}

Before extrapolating to the continuum limit and the physical point, we
interpolate all data to reference strange quark masses as discussed
before. The data for the three strange quark masses are strongly
correlated because the same stochastic light perambulators were used
for all light-strange observables. As a consequence, the
variance-covariance matrix was sometimes not sufficiently well
estimated such that its inverse was unreliable.
As a result, we resort to performing these fits using uncorrelated $\chi^2$ which results in best fit parameters which describe the data much better.
It should be noted that all statistical covariance is still fully taken into account by the bootstrap procedure and our final statistical errors on all fit parameters are correctly estimated.

\begin{figure}[h]
  \centering
  \begin{subfigure}{0.49\textwidth}
    \includegraphics[width=\textwidth]{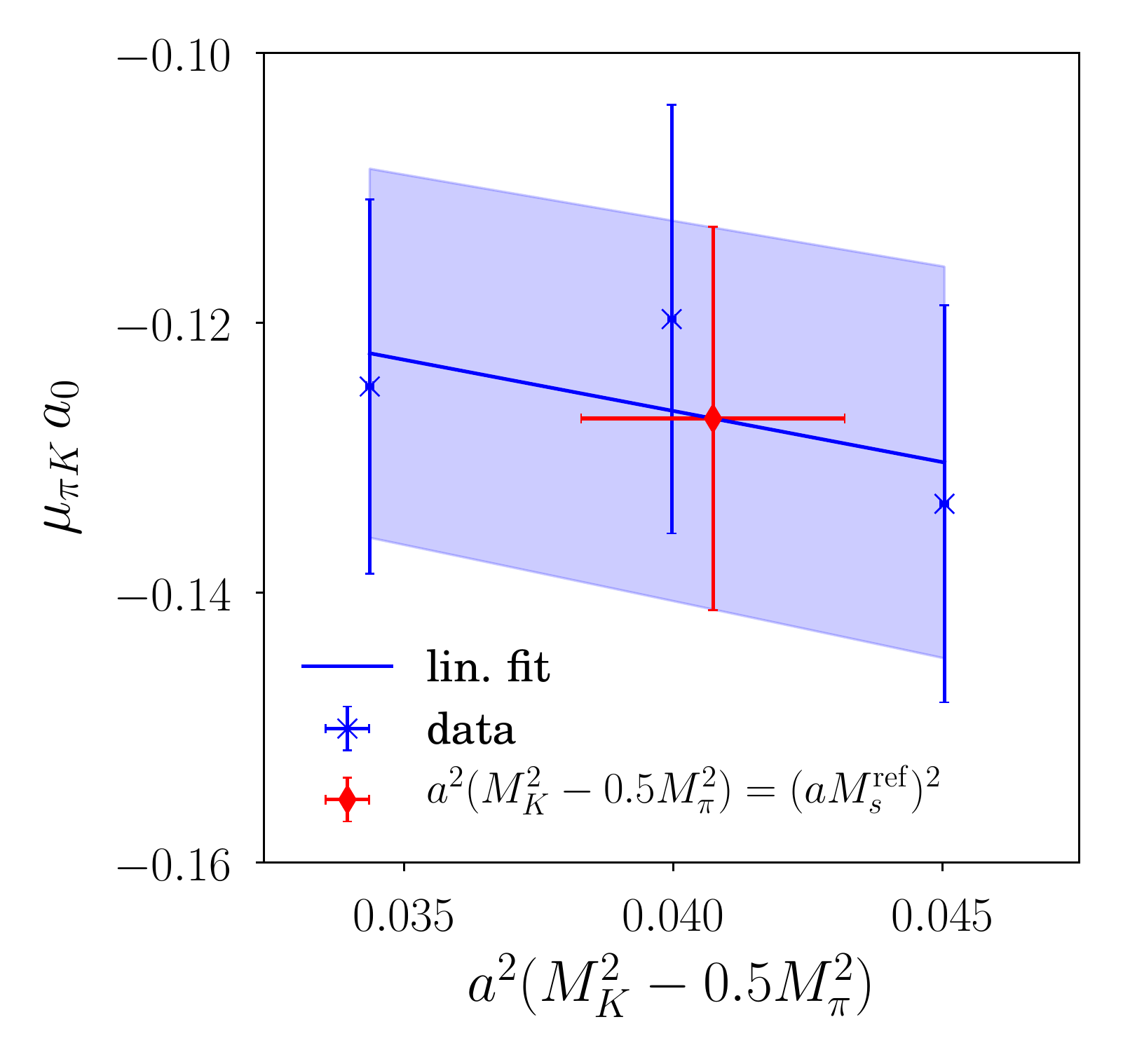}
  \subcaption{B55.32 for \E{1}}
  \label{fig:ma_interp_M1AE2}
  \end{subfigure}
  \begin{subfigure}{0.49\textwidth}
    \includegraphics[width=\textwidth]{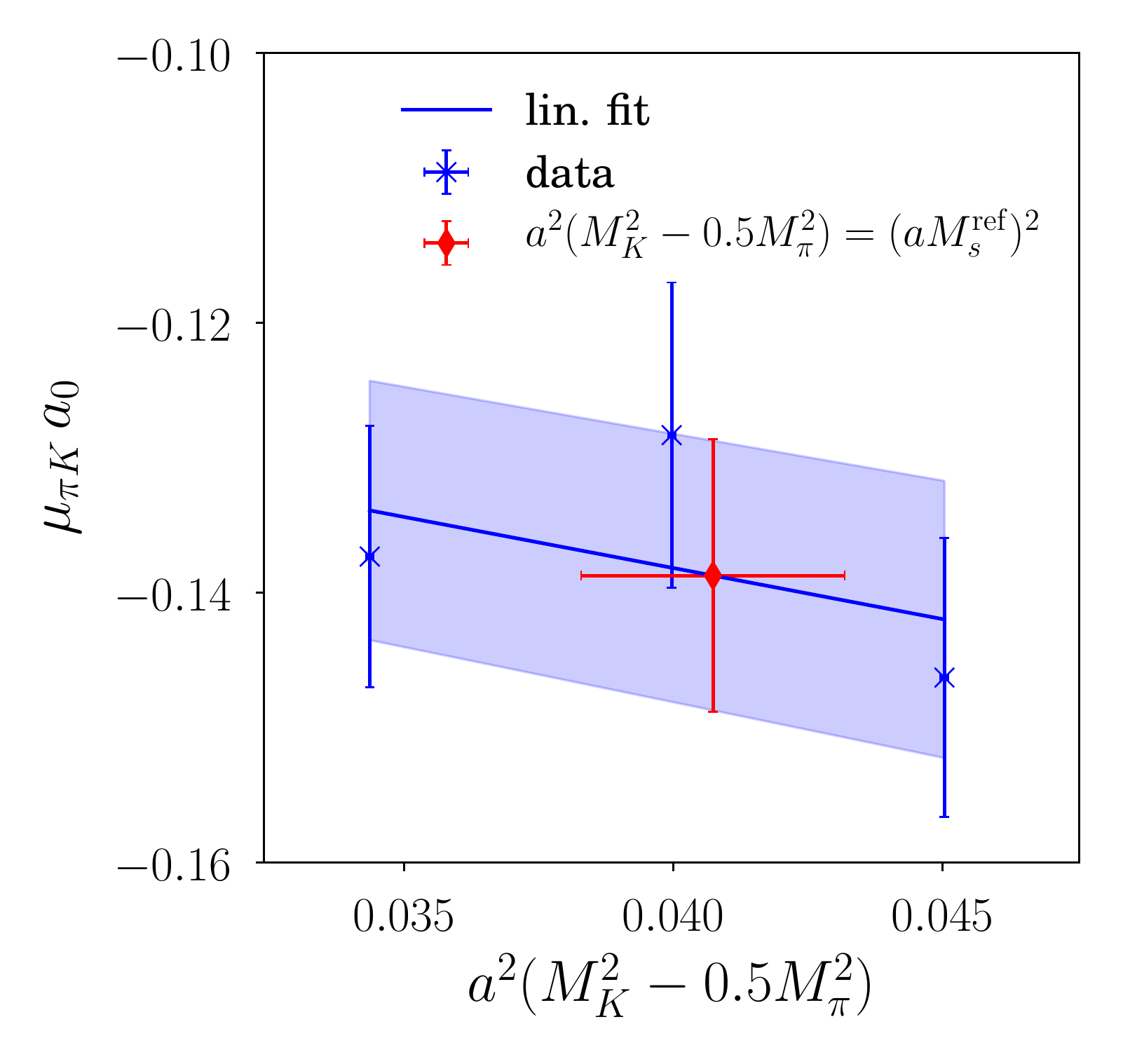}
  \caption{B55.32 for \E{2} }
  \label{fig:ma_interp_mB}
  \end{subfigure}
  \caption{Interpolations of $\ma^{3/2}$ for the two different methods \E{1} and
  \E{2}.}
  \label{fig:ma_interp}
\end{figure}

As an example, we show in \Cref{fig:ma_interp}
the interpolation of $\ma^{3/2}$ in $M_K^2 - 0.5M_\pi^2$ for
ensemble B55.32 comparing methods \E{1} and \E{2}. The
large uncertainty in the interpolation variable stems mainly from the uncertainty in the scaling
quantity $r_0$. Furthermore the errors of the three data points are highly 
correlated.
The interpolation to the reference point is shown as a red
diamond. In general, the strange quark mass dependence of $\ma^{3/2}$ is mild and stems mainly
from the reduced mass $\mu_{\pi K}$.
The values thus determined are compiled in
\Cref{sec:chpt_A_input}. They serve as input data for the subsequent chiral
extrapolations.

\subsection{Chiral and Continuum Extrapolations}
\label{sec:interp}

Having interpolated all our lattice data to a fixed reference strange quark mass corresponding to the physical strange quark mass at leading order, we will describe below possible systematic errors in our chiral and continuum extrapolations.

\subsubsection{\boldmath $\Gamma$-Method}
\label{sec:res_gamma}
\begin{figure}
  \begin{subfigure}{0.49\textwidth}
    \includegraphics{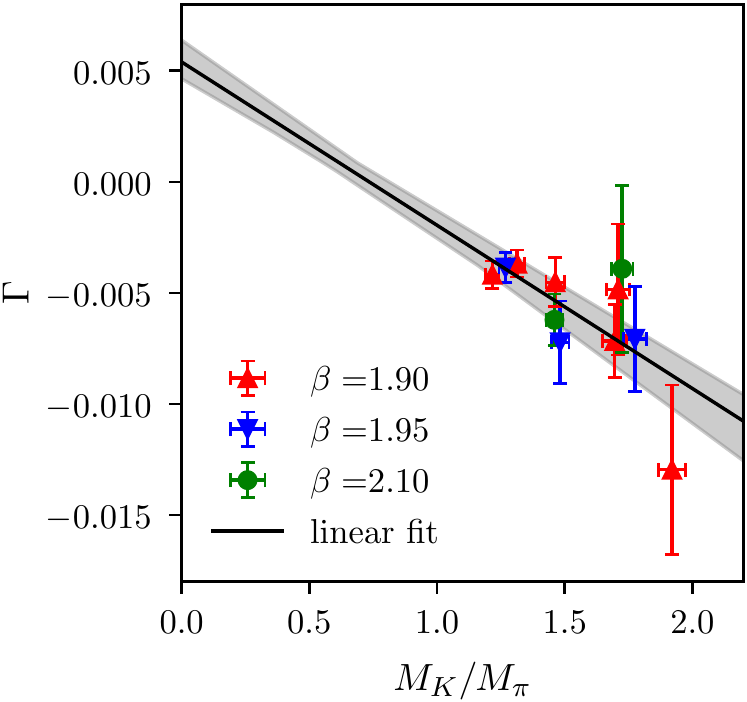}
    \caption{Pollution Removal \E{1}}
  \end{subfigure}
  \begin{subfigure}{0.49\textwidth}
    \includegraphics{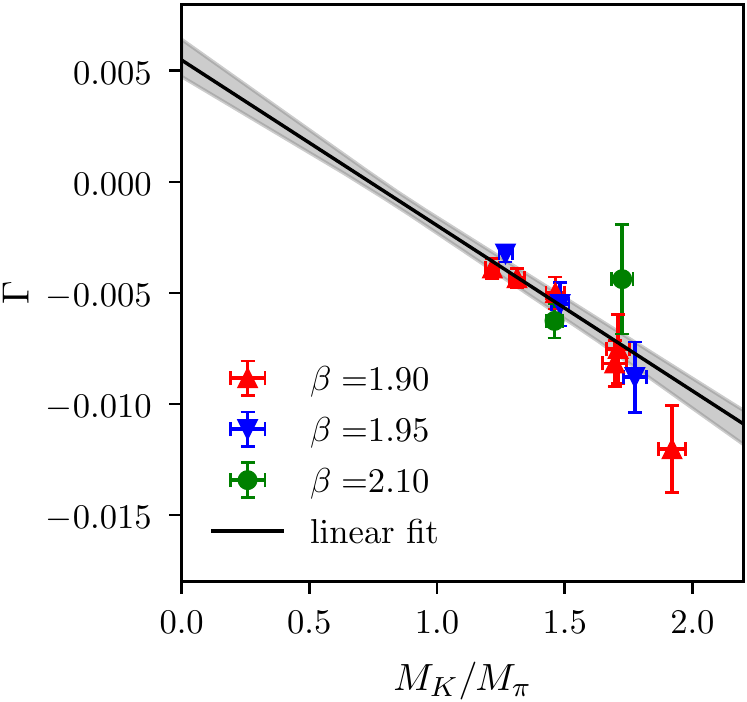}
    \caption{Pollution Removal \E{2}}
  \end{subfigure}
  \caption{Linearized chiral extrapolations of $\Gamma$ with data
    interpolated to the reference strange quark mass. The data for different
    lattice spacings are color encoded. In addition we show the linear fit
    (solid curve, gray error band)}
  \label{fig:lin_chpt_a}
\end{figure}

In this section we present results employing the determination of $L_{\pi K}$ using the linear fit introduced in \Cref{sec:lin_chpt}.
\Cref{fig:lin_chpt_a} shows the chiral extrapolations in terms of $M_K/M_{\pi}$
for pollution removal \E{1} and \E{2}.
Since we work at fixed strange quark mass, the light quark mass decreases from left to right in the figure.
In order to check how our extraction of $L_{\pi K}$ is affected by the range of
included pion masses we employ three different fit ranges
\begin{align}
  \frac{M_K}{M_{\pi}}\in\Set{(0;2.0),(1.5;2.0),(1.4;2.0)}\,.
  \label{eq:fit_intervals_gamma}
\end{align}
In \Cref{tab:lin_chpt_res_ex} we compile the results for the fits corresponding
to the data points of \Cref{fig:lin_chpt_a} for all 3 fit ranges. As the fit
range is restricted to our lightest ensembles, the value extracted
for $L_{\pi K}$ tends up, while the absolute value of the extracted
$\muazerothreehalves$ decreases. It is worth noting that this behavior is only
observed for the pollution removal \E{2}, whereas for \E{1} the values for
$L_{\pi K}$ and $\muazerothreehalves$ stay constant within their statistical
errors.

\begin{table}
  \centering
  \begin{tabular*}{.9\textwidth}{@{\extracolsep{\fill}}l
      c
			c
      S[fixed-exponent=-3,table-omit-exponent]
      S[fixed-exponent=-2,table-omit-exponent]} 
    \hline\hline
    Removal & {Fit range} & $p$-value & {$L_{\pi K}\times 10^{3}$} & {$\muazerothreehalves \times 10^{2}$}\\
    \hline
    \multirow{3}{*}{\E{1}}  & \numrange{0.0}{2.0}&0.8&0.0037 +- 0.0002  &-0.0468 +- 0.0009 \\
                            & \numrange{1.4}{2.0}&0.7&0.0037 +- 0.0002 &   -0.047 +- 0.001 \\
                            & \numrange{1.5}{2.0}&0.6&0.0036 +- 0.0003 &   -0.047 +- 0.002 \\
    \hline
    \multirow{3}{*}{\E{2}}  & \numrange{0.0}{2.0}&0.3&0.0037 +- 0.0001   &-0.0465 +- 0.0007\\
                            & \numrange{1.4}{2.0}&0.5&0.0038 +- 0.0002   &-0.0457 +- 0.0009\\
                            & \numrange{1.5}{2.0}&0.4&0.0040 +- 0.0002  &   -0.045 +- 0.001\\
    \hline\hline
  \end{tabular*}
  \caption{Fit results for the Gamma method. The fits shown in
    \Cref{fig:lin_chpt_a} correspond to the largest fit range in the table.}
  \label{tab:lin_chpt_res_ex}
\end{table}

\subsubsection{Chiral Perturbation Theory at NLO}
\label{sec:i32_chpt}

\begin{figure}[h!]
  \begin{subfigure}{0.49\textwidth}
    \includegraphics{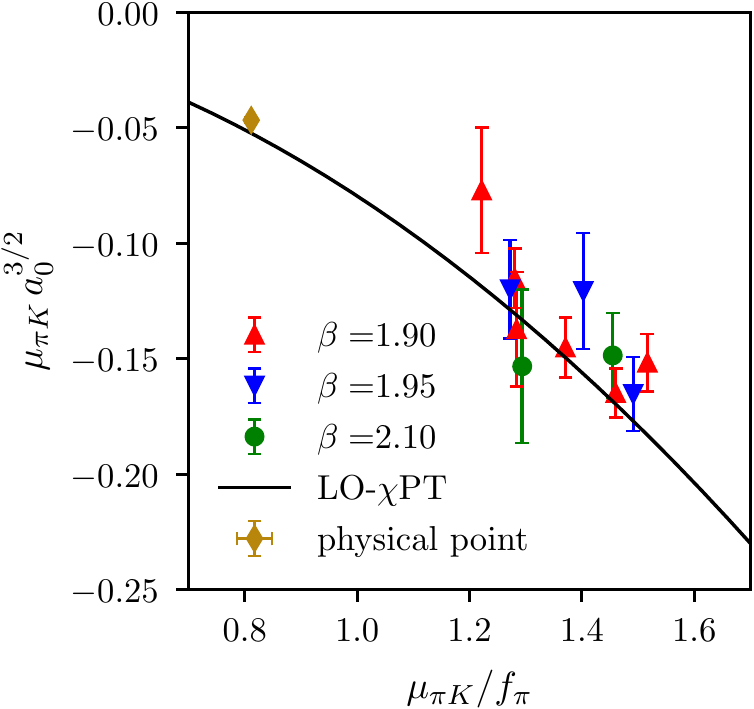}
    \caption{\E{1}:$\ma^{3/2}$ in terms of $\mu_{\pi K}/f_{\pi}$ together with LO ChPT formula.}
    \label{fig:pik_chpt_a1e1}
  \end{subfigure}
  \begin{subfigure}{0.49\textwidth}
    \includegraphics{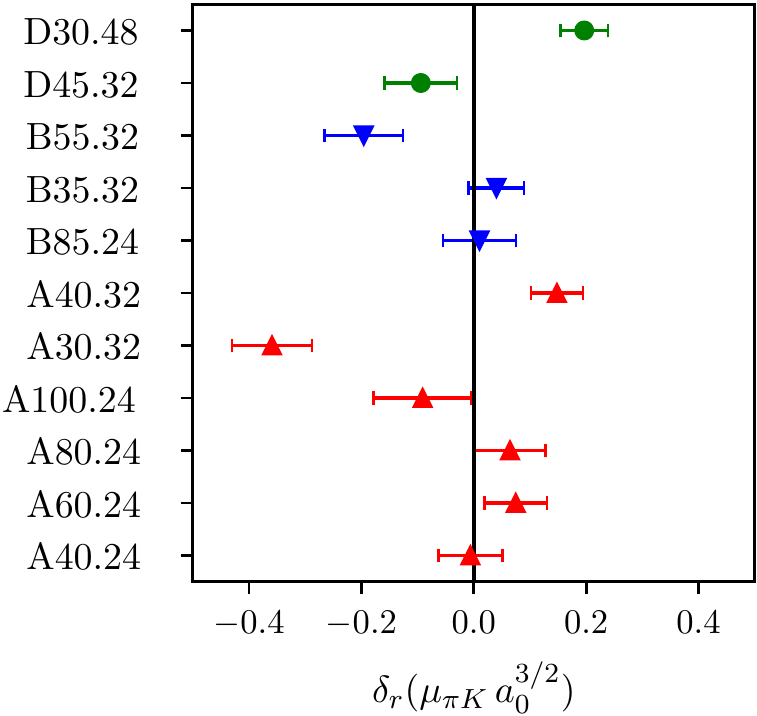}
    \caption{\E{1}:Relative deviation between the measured and the fitted values
    of $\ma^{3/2}$.}
    \label{fig:pik_chpt_a1e1_rel_dev}
  \end{subfigure}
  \begin{subfigure}{0.49\textwidth}
    \includegraphics{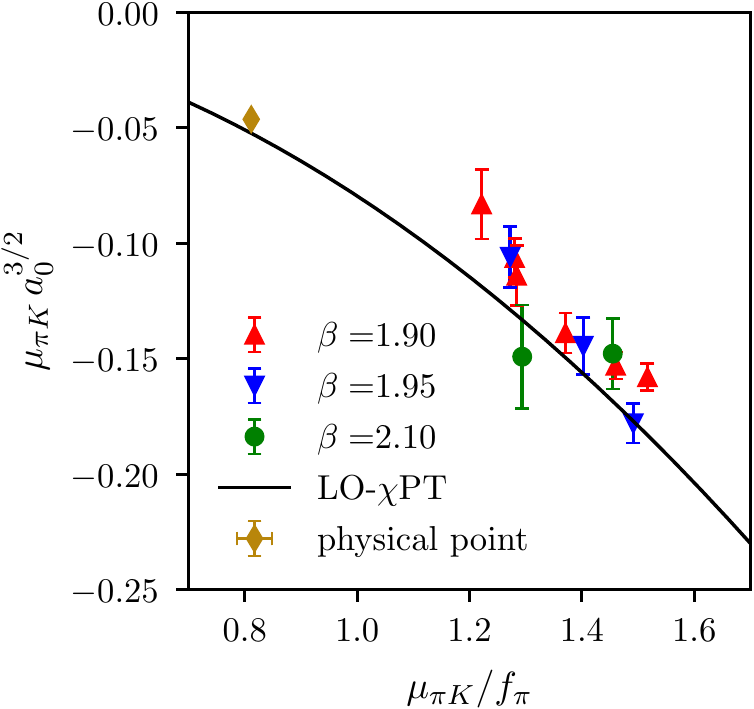}
    \caption{\E{2}:$\ma^{3/2}$ in terms of $\mu_{\pi K}/f_{\pi}$ together with LO ChPT formula.}
    \label{fig:pik_chpt_a1e2}
  \end{subfigure}
  \begin{subfigure}{0.49\textwidth}
    \includegraphics{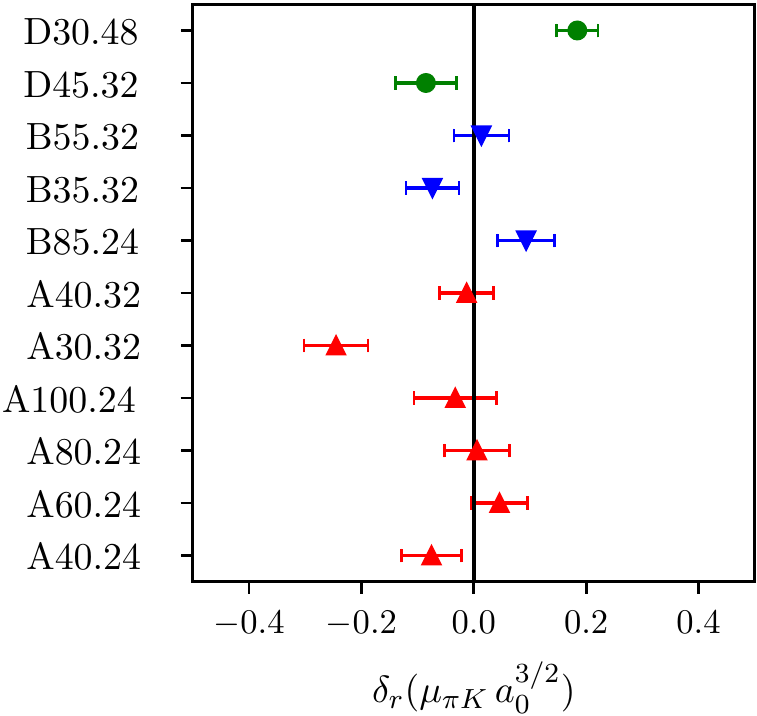}
    \caption{\E{2}:Relative deviation between the measured and the fitted values
    of $\ma^{3/2}$.}
    \label{fig:pik_chpt_a1e2_rel_dev}
   \end{subfigure}
   \caption{Chiral extrapolation of $\ma^{3/2}$ for \E{1} and \E{2}. Different colors and symbols denote different values of $\beta$. In black we plot the LO ChPT formula. The golden diamond gives our final result using the given method at the physical point.}
   \label{fig:pik_chpt_a1}
\end{figure}

We first fit the continuum ChPT formula \Cref{eq:a0_alg_fit} with
$c=0$, i.e. assuming no lattice artefacts.
In the right column of plots in figure \Cref{fig:pik_chpt_a1}, we show
the lattice data for $\ma^{3/2}$ interpolated to the reference strange quark
mass as a function of $\mu_{\pi K}/f_{\pi}$ for the two thermal
pollution removal methods 
\E{1} and \E{2}. The solid line corresponds to the leading order,
parameter free ChPT prediction. Plotting our best fit curve with NLO
ChPT together with the data is difficult, because $\ma^{3/2}$ depends on
meson masses and $f_\pi$ besides $\mu_{\pi K}/f_{\pi}$. Therefore, in
order to demonstrate that the fit is able to describe our data, we
indicate the relative deviation $\delta_r(\ma^{3/2})$ between the fitted
points and the original data 
\begin{align}
  \delta_r(\ma^{3/2})=
  \frac{(\ma^{3/2})^{\text{meas}}-(\ma^{3/2})^{\text{fit}}}
  {(\ma^{3/2})^{\text{meas}}}\,,
       \label{eq:rel_dev_ma}
\end{align}
in \Cref{fig:pik_chpt_a1e1_rel_dev,fig:pik_chpt_a1e2_rel_dev}.
The indicated error bars are statistical only and it is clear that
within these uncertainties, our data is reasonably well described by
the fit. 

As in \Cref{sec:res_gamma}, to investigate the validity of \Cref{eq:a0_alg_fit} across our entire range of pion masses, we studied three different fit intervals for $\mu_{\pi K}/f_{\pi}$, namely
\begin{align}
  \frac{\mu_{\pi K}}{f_{\pi}}\in\Set{(0;1.35),(0;1.41),(0;1.60)}\,,
\end{align}
where now the first range corresponds to only using our lightest pion masses and the third range includes all of our ensembles.
The resulting trend in the extracted values of $L_{\pi K}$ and $\muazerothreehalves$ is shown in \Cref{tab:nlo_chpt_res_ex}.
Just as in the study of the $\Gamma$-method, including heavier pion masses leads
to smaller values of $L_{\pi K}$ and correspondingly smaller values of $\muazerothreehalves$.
\begin{table}
	\centering
	\begin{tabular*}{.9\textwidth}{@{\extracolsep{\fill}}l
		c
		c
		S[fixed-exponent=-3,table-omit-exponent]
		S[fixed-exponent=-2,table-omit-exponent]} 
		\hline\hline
		Removal & {Fit range} & $p$-value & {$L_{\pi K}\times 10^{3}$}  & {$\muazerothreehalves \times 10^{2}$}\\
		\hline
		\multirow{3}{*}{\E{1}}	&\numrange{0.0}{1.35}	&0.6&	0.0036 +- 0.0003 &	-0.047 +- 0.002\\
		                      	&\numrange{0.0}{1.41}	&0.7&	0.0036 +- 0.0002 &	-0.047 +- 0.001\\
		                      	&\numrange{0.0}{1.60}	&0.7&	0.0037 +- 0.0002 &	-0.047 +- 0.001\\
		\hline
		\multirow{3}{*}{\E{2}}	&\numrange{0.0}{1.35}	&0.5&0.0039 +- 0.0002 &	-0.045 +- 0.001 \\
		                   			&\numrange{0.0}{1.41}	&0.5&0.0038 +- 0.0002 &	-0.046 +- 0.001 \\
		                   			&\numrange{0.0}{1.60}	&0.4&0.0037 +- 0.0001 &	-0.0464 +- 0.0007 \\
		\hline\hline
	\end{tabular*}
	\caption{Fit results for the NLO ChPT fit. The fits shown in
	\Cref{fig:pik_chpt_a1} correspond to the largest fit range in the table.}
	\label{tab:nlo_chpt_res_ex}
\end{table}

To investigate possible discretization effects we now allow $c\neq 0$ in \Cref{eq:a0_alg_fit}. 
Fitting \Cref{eq:a0_alg_fit} for the different choices of $f(a^2)$, we are neither able to obtain a statistically significant result for the fit parameter $c$, nor do we see significant differences in the extracted values of $L_{\pi K}$ and $\muazerothreehalves$.
We conclude that we are not able to resolve lattice artifacts in this
quantity given our statistical uncertainties.

 \section{Discussion}
\label{sec:discussion}

Let us first discuss the main systematics of our computation:
In contrast to the pion-pion or kaon-kaon systems, there are time
dependent thermal pollutions in the correlation functions relevant for the
extraction of the pion-kaon s-wave scattering length. This very fact
turns out to represent one of the major systematic uncertainties in
the present computation. We have investigated two methods to
remove the leading thermal pollutions, denoted as \E{1} and \E{2}.
With both we are able to describe the data for the correlation
functions. However, there is uncertainty left, because we remove only
the leading pollution and the removal procedure requires input
estimated from other two point functions. Thus, we eventually decided
to use both methods \E{1} and \E{2} and include the differences in the
systematic uncertainty. 

Secondly, we perform a mixed action simulation for the strange quark. We
use this to correct for small mis-tuning in the sea strange quark mass
value used for the gauge configuration generation. This leads --- at
least in principle --- to a small mismatch in the renormalization
condition used for the continuum extrapolation. We cannot resolve the
corresponding effect on our results quantitatively given our
statistical uncertainties. But, since we
study quantities which mainly depend on the valence quark
properties we expect them to be small. 

Thirdly, in the ChPT determination of $L_{\pi K}$ the remaining LEC, $L_5$,
entered as a prior to numerically stabilize our fits. The HPQCD value of
Ref.~\cite{Dowdall:2013rya} stems from an independent lattice
simulation, but is extrapolated to the continuum limit.
In Ref.~\cite{Dowdall:2013rya} $L_5$ is given at scale $M_\eta$,
which we translated to our renormalization scale given by the pion
decay constant.

In addition, the extrapolation from our data to the physical point is
quite long. Here, a computation directly with physical pion mass would
improve our confidence in the result. The final error on our
determination is only as small as it is due to the highly constraining 
ChPT description of $\muazerothreehalves$.

Finally, although we are not able to resolve lattice artefacts in
our determination of $\ma^{3/2}$ our statistical errors and limited set
of gauge ensembles especially at the finest lattice spacing might
make us unable to resolve possible lattice artefacts.

\section{Summary}
\label{sec:summary}

\begin{figure}[t!]
  \centering
  \begin{subfigure}{.49\textwidth}
    \includegraphics[height=210pt]{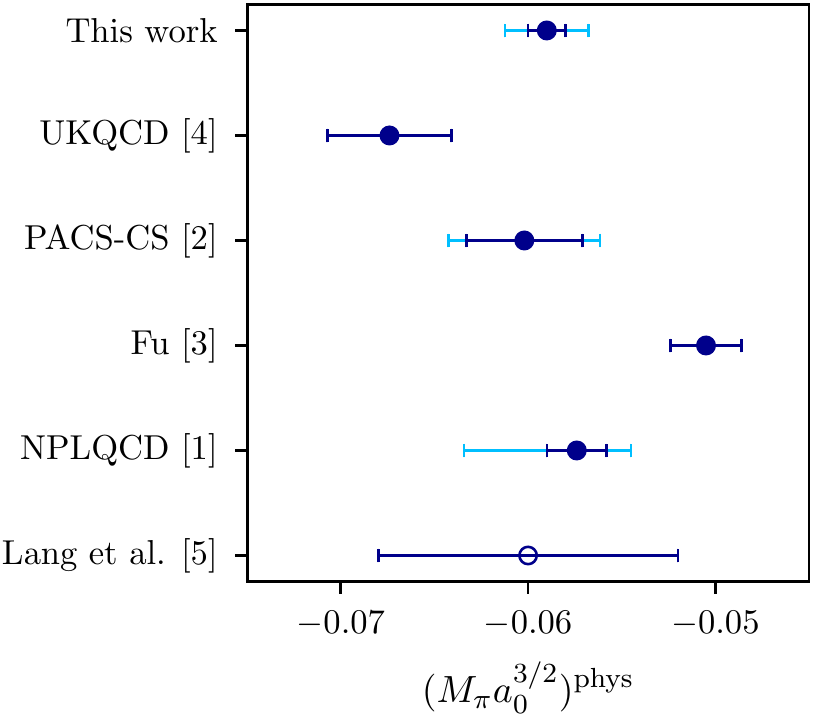}
  \end{subfigure}
  \begin{subfigure}{.49\textwidth}
    \includegraphics[height=210pt]{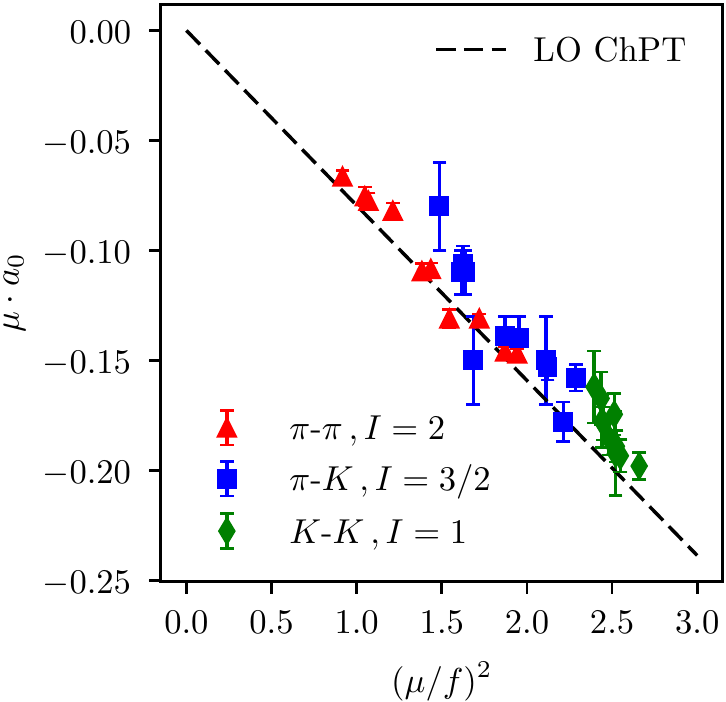}
  \end{subfigure}
  \caption{Left: Comparison of physical results for $M_\pi a_0^{I=3/2}$
    from various lattice
    computations~\cite{Beane:2006gj,Sasaki:2013vxa,Fu:2011wc,Janowski:2014uda,Lang:2012sv}.
    The unfilled point denotes the LO extrapolation to the physical point using
    the data of Ref.~\cite{Lang:2012sv}.
    Right: s-wave scattering lengths for the pion-pion, pion-kaon and
    kaon-kaon maximum isospin channels as a function of the squared
    reduced mass $\mu^2$ of the system divided by $f_\pi^2$ for pion-pion and
    pion-kaon and by $f_K^2$ for kaon-kaon.}
  \label{fig:GBsummary}
\end{figure}

In this paper we have presented a first lattice computation of the
pion-kaon s-wave scattering length for isospin $I=3/2$ extrapolated to
the continuum limit. By varying our methodology we estimate the 
systematic uncertainties in our results. Our errors cover statistical
uncertainties, continuum and chiral extrapolations as well as the
removal of thermal pollutions.

In the left panel of \Cref{fig:GBsummary} we compare the results
presented in this paper with previous lattice determinations. The
inner (darker)
error bars show the purely statistical errors whereas the outer (lighter) ones correspond
to the statistical and systematic errors added in quadrature. 
Even though the four other determinations lack the extrapolation to the continuum
limit, overall agreement within errors is observed. However,
concerning the final uncertainty, our determination improves
significantly on the previous determinations by controlling more
sources of uncertainty. 

As mentioned in the introduction, the three two particle systems
pion-pion, pion-kaon and kaon-kaon are very similar. Therefore, it is
interesting to compare the data for pion-pion~\cite{Helmes:2015gla},
kaon-kaon~\cite{Helmes:2017smr} and pion-kaon in a single plot.
This is done in the right panel of \Cref{fig:GBsummary} where we show $\mu\cdot a_0$ as a
function of $(\mu/f)^2$. Here $\mu$ is the reduced mass of the
corresponding two particle system and $f$ is the pion decay constant
$f_\pi$ for the pion-pion and pion-kaon and $f_K$ for
the kaon-kaon system. The dashed line in the right panel of \Cref{fig:GBsummary}
is the leading order, parameter-free ChPT prediction all three systems share. The
three symbols (and colors) represent our data for the three different
systems, respectively.

It can be seen that for all three systems, the deviations from LO ChPT are small. 
For the pion-kaon system, a parametrization in terms of $f_K \cdot f_\pi$ would bring the points even closer to the LO line, while increasing the deviation of the final result from the LO estimate.
For the kaon-kaon system, instead, a parametrization in terms of $f_\pi$ rather than $f_K$ (which is perfectly valid at this order of ChPT) would render the deviation from the LO line more severe.

It is somewhat surprising that ChPT appears to work so well for all
three systems, especially for the heavier points in our simulations
and even more for the kaon-kaon system,
where the expansion parameter becomes large. A possible reason for
this finding might be the fact that all three systems are only weakly
interacting. 

\section*{Acknowledgements}

We thank the
members of ETMC for the most enjoyable collaboration. The computer
time for this project was made available to us by the John von
Neumann-Institute for Computing (NIC) on the Juqueen and Jureca
systems in J{\"u}lich. 
We thank K.~Ottnad for providing us with the data for $f_\pi$ and S.~Simula for the estimates of the finite size
corrections to $M_\pi$, $M_K$ and $f_\pi$.
This project was funded by the DFG as a project in
the Sino-German CRC110. The open source software
packages tmLQCD~\cite{Jansen:2009xp}, Lemon~\cite{Deuzeman:2011wz}, QUDA~\cite{Clark:2009wm,Babich:2011np,Clark:2016rdz},
R~\cite{R:2005}, python~\cite{python:2001} and SciPy~\cite{scipy:2018} have been used.
 \bibliographystyle{h-physrev5}

\begin{appendix}
  \crefalias{section}{appsec}
  \crefalias{subsection}{appsec}
  \section{\boldmath $\pi$-$K$ Scattering in ChPT}
\label{app:pik_chpt}
\subsection{Isospin even/odd Scattering Amplitudes}
For completeness, we reproduce here the derivation of the $\Gamma$ method~\cite{Beane:2006gj} described in \Cref{sec:lin_chpt}.
The scattering amplitudes for all isospin channels of $\pi$-$K$ scattering can
be noted down using basis elements that are even (odd) under exchange of the
Mandelstam variables $s$ and $u$ 
\begin{align}
	\mathcal{A}^+ = \frac{1}{3}\left(\mathcal{A}^{1/2}(s,t,u)+2\mathcal{A}^{3/2}(s,t,u)\right)\,\nonumber\\
	\mathcal{A}^- = \frac{1}{3}\left(\mathcal{A}^{1/2}(s,t,u)-\mathcal{A}^{3/2}(s,t,u)\right)\,.
	\label{eq:iso_eo}
\end{align}
For those amplitudes expressions for the scattering lengths $a^+$ and $a^-$ can
be derived. They are noted down, for example, in
Ref.~\cite{Kubis:2001ij}. Note that
$a^+$ only depends on $L_5$ while $a^-$ only depends on $L_{\pi K}$.
From \Cref{eq:iso_eo} it follows that
\begin{align}
	\mathcal{A}^{1/2} = \mathcal{A}^+ + 2\mathcal{A}^- \\
	\mathcal{A}^{3/2} = \mathcal{A}^+ - \mathcal{A}^-\,,
	\label{eq:def_iso}
\end{align}
which immediately carries over to the scattering lengths $a^{1/2}$ and
$a^{3/2}$ yielding
\begin{align}
	\muazeroonehalf = \mu_{\pi K}(a^++2a^-)\\
	\muazerothreehalves = \mu_{\pi K}(a^+-a^-)\,.
	\label{eq:def_iso_scat}
\end{align}
Inserting the ChPT formulae for $a^+$ and $a^-$ into \Cref{eq:def_iso_scat} one
obtains the ChPT formulae \Cref{eq:a0_alg_fit}. Moreover, after the insertion of
$a^+$ and $a^-$ into \Cref{eq:def_iso_scat} the terms containing the LECs $L_5$
and $L_{\pi K}$ can be collected at one side in order to
arrive at \Cref{eq:gamma_fit}. The expression $\Gamma$ in \Cref{eq:gamma_fit} is
then given by \Cref{eq:gamma_calc}
\subsection{Next to Leading Order Functions}
\label{app:nlo}
For convenience we list the chiral functions $\chi^{\pm}_{\text{NLO}}$ derived
in Ref.~\cite{Kubis:2001ij}
\begin{align}
	\chi^{+}_{\text{NLO}}(\Lambda_\chi,M_\pi,M_K) = \frac{1}{16\pi^2}&\left[
	\nu_\pi \chilog{M_\pi} + \nu_K\chilog{M_K} +
	\nu_\eta \chilog{M_\eta} \right.\nonumber \\
	&\left.+\nu_{\tan} \arctan\left( \frac{2(M_K + M_\pi )}{M_K - 2M_\pi }
	\sqrt{ \frac{M_K - M_\pi }{ 2M_K + M_\pi }}\right)\right.\nonumber\\
	&\left.+\nu{'}_{\tan} \arctan\left( \frac{2(M_K - M_\pi )}{M_K + 2M_\pi}
	\sqrt{ \frac{M_K +M_\pi }{ 2M_K - M_\pi }} \right)
+\frac{43}{9}\right]\,,\\
	\chi^{-}_{\text{NLO}}(\Lambda_\chi,M_\pi,M_K) = \frac{M^2_\pi }{8f^2_\pi \pi^2}&\left[
	\nu{'}_\pi \chilog{M_\pi} + \nu{'}_K\chilog{M_K} + 
	\nu{'}_\eta \chilog{M_\eta} \right.\nonumber \\
	&\left.+\frac{M_K}{M_\pi}\nu{'}_{\tan}
	\arctan\left( \frac{2(M_K - M_\pi)}{M_K + 2M_\pi}
	\sqrt{ \frac{M_K +M_\pi }{ 2M_K - M_\pi }}\right) \right]\,.
	\label{eq:nlo_eo}
\end{align}
The functions $\nu^{(\prime)}_X$ are given by
\begin{align}
	\nu_\pi &= \frac{11M^2_\pi }{2(M^2_K-M^2_\pi )} \\
	\nu_K &= -\frac{67M^2_K-8M^2_\pi }{9(M^2_K-M^2_\pi )} \\
	\nu_\eta &= +\frac{24M^2_K-5M^2_\pi }{18(M^2_K-M^2_\pi )} \\
	\nu_{\tan} &= -\frac{4}{9}\frac{\sqrt{2M^2_K -M_K M_\pi -M^2_\pi }}{M_K +M_\pi
	} \\
	\nu{'}_\pi &= -\frac{8M^2_\pi -5M^2_\pi }{2(M^2_K-M^2_\pi )} \\
	\nu{'}_K &= \frac{23M^2_K}{9(M^2_K-M^2_\pi )} \\
	\nu{'}_\eta &= \frac{28M^2_K-9M^2_\pi }{18(M^2_K-M^2_\pi )} \\
	\nu{'}_{\tan} &=-\frac{4}{9}\frac{\sqrt{2M_K^2 -M_K M_\pi -M^2_\pi }}{M_K
	+M_\pi }\,.
\end{align}
From these isospin even/odd functions the definite isospin
functions $\chi^{3/2}_{\text{NLO}}$ and $\chi^{1/2}_{\text{NLO}}$ can be
derived in the same way as the scattering lengths of \Cref{eq:def_iso_scat}
\begin{align}
  &+ \kappa_{\tan}^{\prime} \arctan
  \left( \frac{2(M_K + M_\pi)}{M_K - 2M_\pi}
  \sqrt{ \frac{M_K - M_\pi }{ 2M_K + M_\pi }} \right) \,.	
  \label{eq:nlo_def_iso}
\end{align}
Here the functions $\kappa^{(\prime)}_X$ are given by
\begin{align}
\kappa_\pi &= \frac{11M_KM_\pi^3 + 8M_\pi^2M_K^2 -5M_\pi^4}{2(M_K^2 -M_\pi^2)}\\
\kappa_K   &= -\frac{67M_K^3 M_\pi - 8M_\pi^3M_K + 23M_K^2M_\pi^2}{9(M_K^2-M_\pi^2)}\\
\kappa_\eta &= \frac{24M_\pi M_K^3 -5 M_K M_\pi^3 + 28M_K^2 M_\pi^2 - 9M_\pi^4}
     {18(M_K^2-M_\pi^2)} \\
\kappa_{\tan} &= -\frac{16 M_KM_\pi}{9}\frac{\sqrt{2M_K^2 +M_KM_\pi -M_\pi^2}}{M_K-M_\pi} \\
\kappa_{\pi}^{'}  &= \frac{11M_KM_\pi^3-16M_K^2M_\pi^2+10M_\pi^4}{2(M_K^2-M_\pi^2)} \\
\kappa_{K}^{'}    &=  - \frac{67 M_{K}^3 M_\pi-8 M_{\pi}^3 M_K-46 M_{K}^2 M_{\pi}^2}{9(M_K^2-M_\pi^2)} \\
\kappa_{\eta}^{'} &= \frac{24M_\pi M_K^3-5M_KM_\pi^3-56M_K^2M_\pi^2+18M_\pi^4}{18(M_K^2-M_\pi^2)} \\
\kappa_{\tan}^{'}  &= \frac{8 M_K M_\pi}{9}\frac{\sqrt{2M_K^2 -M_K M_\pi -M_\pi^2}}{M_K +M_\pi} \,.
\end{align}

\section{Datatables}
\begin{table}
	\centering
	\begin{tabular*}{.9\textwidth}{
      @{\extracolsep{\fill}}
      l@{}
		S[round-mode=places,round-precision=1]@{} 
		S[round-mode=places,round-precision=1]@{} 
		S[round-mode=figures,round-precision=1]@{} 
		S[round-mode=figures,round-precision=5]} 
		\hline\hline
		Ensemble & {$f_\pi$} & {$K^{\text{FSE}}_{f_{\pi}}$} & {$K^{\text{FSE}}_{M_{\pi}}$} & {$K^{\text{FSE}}_{M_K}$} \\
		\hline
   		A30.32 & 0.06452 \pm 0.00021 & 0.9757 \pm 0.0061& 1.0081 \pm 0.0052 & 1.002327 \pm 0.000001\\ 
   		A40.24 & 0.06577 \pm 0.00024 & 0.9406 \pm 0.0084& 1.0206 \pm 0.0095 & 1.009874 \pm 0.000001\\
   		A40.32 & 0.06839 \pm 0.00018 & 0.9874 \pm 0.0024& 1.0039 \pm 0.0028 & 1.001299 \pm 0.000001\\
   		A60.24 & 0.07209 \pm 0.00020 & 0.9716 \pm 0.0037& 1.0099 \pm 0.0049 & 1.004681 \pm 0.000001\\
   		A80.24 & 0.07581 \pm 0.00013 & 0.9839 \pm 0.0022& 1.0057 \pm 0.0029 & 1.002518 \pm 0.000001\\
   		A100.24 & 0.07936 \pm 0.00014 & 0.9900 \pm 0.0015& 1.0037 \pm 0.0019 & 1.001480\pm 0.000001\\
   		B35.32 & 0.06105 \pm 0.00017 & 0.9794 \pm 0.0027& 1.0069 \pm 0.0032 & 1.002466 \pm 0.000001\\
   		B55.32 & 0.06545 \pm 0.00011 & 0.9920 \pm 0.0010& 1.0027 \pm 0.0014 & 1.000879 \pm 0.000001\\
   		B85.24 & 0.07039 \pm 0.00026 & 0.9795 \pm 0.0024& 1.0083 \pm 0.0028 & 1.003178 \pm 0.000001\\
   		D30.48 & 0.04735 \pm 0.00015 &  0.9938\pm 0.0005& 1.0021 \pm 0.0007 & 1.000714 \pm 0.000001\\
   		D45.32 & 0.04825 \pm 0.00014 & 0.9860 \pm 0.0013& 1.0047 \pm 0.0014 & 1.000000 \pm 0.000001\\
		\hline\hline
	\end{tabular*}
	\caption{External data used via parametric bootstrapping. The error on $K^{\text{FSE}}_{M_K}$ is only estimated}
	\label{tab:ext_dat}
\end{table}

\begin{table}
  \centering
  \begin{tabular*}{.9\textwidth}{@{\extracolsep{\fill}}lllll}
  \hline\hline
Ensemble &  $a\mu_{s}$ & $aM_{K}$ & $aM_{\pi}$ &  $a\mu_{\pi K}$\\
\hline
\multirow{3}{*}{A30.32} & 0.0185 & $0.2294(3)(^{+0}_{-0})$ & $0.1239(2)(^{+1}_{-0})$ & $0.08046(12)(^{+0}_{-0})$ \\
          & 0.0225 & $0.2495(2)(^{+1}_{-0})$ & $0.1239(2)(^{+1}_{-0})$ & $0.08280(12)(^{+0}_{-0})$ \\
          & 0.0246 & $0.2597(2)(^{+1}_{-0})$ & $0.1239(2)(^{+1}_{-0})$ & $0.08388(12)(^{+0}_{-0})$ \\
\hline
\multirow{3}{*}{A40.24} & 0.0185 & $0.2365(5)(^{+2}_{-1})$ & $0.1453(5)(^{+2}_{-1})$ & $0.08999(22)(^{+0}_{-0})$ \\
          & 0.0225 & $0.2561(4)(^{+4}_{-1})$ & $0.1453(5)(^{+2}_{-1})$ & $0.09269(22)(^{+0}_{-0})$ \\
          & 0.0246 & $0.2662(5)(^{+1}_{-2})$ & $0.1453(5)(^{+2}_{-1})$ & $0.09398(22)(^{+0}_{-0})$ \\
\hline
\multirow{3}{*}{A40.32} & 0.0185 & $0.2343(2)(^{+0}_{-0})$ & $0.1415(2)(^{+1}_{-0})$ & $0.08822(10)(^{+0}_{-0})$ \\
          & 0.0225 & $0.2538(2)(^{+1}_{-0})$ & $0.1415(2)(^{+1}_{-0})$ & $0.09086(11)(^{+0}_{-0})$ \\
          & 0.0246 & $0.2638(2)(^{+1}_{-1})$ & $0.1415(2)(^{+1}_{-0})$ & $0.09210(11)(^{+0}_{-0})$ \\
\hline
\multirow{3}{*}{A60.24} & 0.0185 & $0.2448(3)(^{+1}_{-0})$ & $0.1729(3)(^{+4}_{-1})$ & $0.10134(16)(^{+0}_{-0})$ \\
          & 0.0225 & $0.2637(3)(^{+1}_{-0})$ & $0.1729(3)(^{+4}_{-1})$ & $0.10445(16)(^{+0}_{-0})$ \\
          & 0.0246 & $0.2735(3)(^{+1}_{-0})$ & $0.1729(3)(^{+4}_{-1})$ & $0.10594(17)(^{+0}_{-0})$ \\
\hline
\multirow{3}{*}{A80.24} & 0.0185 & $0.2548(2)(^{+0}_{-1})$ & $0.1993(2)(^{+0}_{-1})$ & $0.11184(11)(^{+0}_{-0})$ \\
          & 0.0225 & $0.2731(2)(^{+1}_{-1})$ & $0.1993(2)(^{+0}_{-1})$ & $0.11523(11)(^{+0}_{-0})$ \\
          & 0.0246 & $0.2824(2)(^{+2}_{-2})$ & $0.1993(2)(^{+0}_{-1})$ & $0.11685(11)(^{+0}_{-0})$ \\
\hline
\multirow{3}{*}{A100.24} & 0.0185 & $0.2642(2)(^{+0}_{-1})$ & $0.2223(2)(^{+1}_{-1})$ & $0.12073(11)(^{+0}_{-0})$ \\
          & 0.0225 & $0.2822(2)(^{+0}_{-0})$ & $0.2223(2)(^{+1}_{-1})$ & $0.12436(11)(^{+0}_{-0})$ \\
          & 0.0246 & $0.2913(2)(^{+1}_{-1})$ & $0.2223(2)(^{+1}_{-1})$ & $0.12609(11)(^{+0}_{-0})$ \\
\hline
\multirow{3}{*}{B35.32} & 0.0160 & $0.2053(2)(^{+1}_{-1})$ & $0.1249(2)(^{+1}_{-1})$ & $0.07765(11)(^{+0}_{-0})$ \\
          & 0.0186 & $0.2186(2)(^{+2}_{-1})$ & $0.1249(2)(^{+1}_{-1})$ & $0.07948(12)(^{+0}_{-0})$ \\
          & 0.0210 & $0.2298(2)(^{+0}_{-1})$ & $0.1249(2)(^{+1}_{-1})$ & $0.08091(12)(^{+0}_{-0})$ \\
\hline
\multirow{3}{*}{B55.32} & 0.0160 & $0.2155(2)(^{+1}_{-2})$ & $0.1554(2)(^{+0}_{-0})$ & $0.09030(10)(^{+0}_{-0})$ \\
          & 0.0186 & $0.2282(2)(^{+1}_{-2})$ & $0.1554(2)(^{+0}_{-0})$ & $0.09245(10)(^{+0}_{-0})$ \\
          & 0.0210 & $0.2390(2)(^{+1}_{-2})$ & $0.1554(2)(^{+0}_{-0})$ & $0.09418(10)(^{+0}_{-0})$ \\
\hline
\multirow{3}{*}{B85.24} & 0.0160 & $0.2313(3)(^{+0}_{-3})$ & $0.1933(3)(^{+1}_{-0})$ & $0.10530(15)(^{+0}_{-0})$ \\
          & 0.0186 & $0.2429(3)(^{+0}_{-0})$ & $0.1933(3)(^{+1}_{-0})$ & $0.10763(16)(^{+0}_{-0})$ \\
          & 0.0210 & $0.2535(3)(^{+2}_{-0})$ & $0.1933(3)(^{+1}_{-0})$ & $0.10967(15)(^{+0}_{-0})$ \\
\hline
\multirow{3}{*}{D45.32} & 0.0130 & $0.1658(3)(^{+1}_{-1})$ & $0.1205(4)(^{+1}_{-1})$ & $0.06979(17)(^{+0}_{-0})$ \\
          & 0.0150 & $0.1747(4)(^{+4}_{-1})$ & $0.1205(4)(^{+1}_{-1})$ & $0.07132(17)(^{+0}_{-0})$ \\
          & 0.0180 & $0.1876(3)(^{+0}_{-1})$ & $0.1205(4)(^{+1}_{-1})$ & $0.07339(17)(^{+0}_{-0})$ \\
\hline
\multirow{3}{*}{D30.48} & 0.0115 & $0.1503(1)(^{+0}_{-0})$ & $0.0976(1)(^{+0}_{-0})$ & $0.05917(6)(^{+0}_{-0})$ \\
          & 0.0150 & $0.1673(1)(^{+0}_{-1})$ & $0.0976(1)(^{+0}_{-0})$ & $0.06163(6)(^{+0}_{-0})$ \\
          & 0.0180 & $0.1807(1)(^{+1}_{-1})$ & $0.0976(1)(^{+0}_{-0})$ & $0.06336(6)(^{+0}_{-0})$ \\
\hline\hline 
  \end{tabular*} 
\caption{Comparison of the meson masses, $M_{\pi}$ and $M_K$
													together with the reduced mass $\mu_{\pi K}$. 
													The systematic uncertainties for $\mu_{\pi K}$ turn
													out to be negligible and thus are not shown.}
  \label{tab:raw_data_masses_comp}
\end{table}
\begin{table}
  \centering
  \begin{tabular*}{.9\textwidth}{@{\extracolsep{\fill}}llll}
  \hline\hline
Ensemble & $a\mu_{s}$ & $aE_{\pi K}(\E{1})$ &  $aE_{\pi K}(\E{2})$\\
\hline \hline
\multirow{3}{*}{A30.32} & 0.0185 & $0.3558(9)(^{+7}_{-2})$ & $0.3558(7)(^{+5}_{-1})$ \\
          & 0.0225 & $0.3758(9)(^{+8}_{-2})$ & $0.3758(7)(^{+6}_{-1})$ \\
          & 0.0246 & $0.3870(9)(^{+9}_{-5})$ & $0.3867(6)(^{+2}_{-2})$ \\
\hline
\multirow{3}{*}{A40.24} & 0.0185 & $0.3892(12)(^{+6}_{-0})$ & $0.3887(11)(^{+4}_{-2})$ \\
          & 0.0225 & $0.4081(12)(^{+10}_{-2})$ & $0.4082(11)(^{+5}_{-0})$ \\
          & 0.0246 & $0.4184(12)(^{+7}_{-0})$ & $0.4182(11)(^{+5}_{-0})$ \\
\hline
\multirow{3}{*}{A40.32} & 0.0185 & $0.3793(6)(^{+3}_{-8})$ & $0.3789(5)(^{+0}_{-4})$ \\
          & 0.0225 & $0.3988(6)(^{+9}_{-7})$ & $0.3983(5)(^{+1}_{-3})$ \\
          & 0.0246 & $0.4081(6)(^{+2}_{-4})$ & $0.4083(5)(^{+1}_{-4})$ \\
\hline
\multirow{3}{*}{A60.24} & 0.0185 & $0.4250(9)(^{+11}_{-0})$ & $0.4247(7)(^{+8}_{-0})$ \\
          & 0.0225 & $0.4447(7)(^{+2}_{-0})$ & $0.4440(6)(^{+3}_{-4})$ \\
          & 0.0246 & $0.4537(7)(^{+8}_{-4})$ & $0.4536(6)(^{+5}_{-1})$ \\
\hline
\multirow{3}{*}{A80.24} & 0.0185 & $0.4613(6)(^{+0}_{-2})$ & $0.4606(5)(^{+3}_{-0})$ \\
          & 0.0225 & $0.4789(6)(^{+5}_{-0})$ & $0.4787(5)(^{+4}_{-0})$ \\
          & 0.0246 & $0.4894(6)(^{+0}_{-6})$ & $0.4882(5)(^{+3}_{-0})$ \\
\hline
\multirow{3}{*}{A100.24} & 0.0185 & $0.4921(5)(^{+6}_{-3})$ & $0.4922(4)(^{+3}_{-0})$ \\
          & 0.0225 & $0.5102(5)(^{+5}_{-0})$ & $0.5102(4)(^{+3}_{-0})$ \\
          & 0.0246 & $0.5193(5)(^{+3}_{-3})$ & $0.5193(4)(^{+2}_{-1})$ \\
\hline
\multirow{3}{*}{B35.32} & 0.0160 & $0.3333(9)(^{+11}_{-0})$ & $0.3336(6)(^{+6}_{-0})$ \\
          & 0.0186 & $0.3474(7)(^{+2}_{-1})$ & $0.3472(6)(^{+3}_{-1})$ \\
          & 0.0210 & $0.3595(9)(^{+0}_{-3})$ & $0.3584(7)(^{+5}_{-0})$ \\
\hline
\multirow{3}{*}{B55.32} & 0.0160 & $0.3743(5)(^{+4}_{-1})$ & $0.3747(4)(^{+3}_{-0})$ \\
          & 0.0186 & $0.3866(5)(^{+6}_{-7})$ & $0.3869(4)(^{+3}_{-1})$ \\
          & 0.0210 & $0.3977(5)(^{+4}_{-1})$ & $0.3981(4)(^{+3}_{-0})$ \\
\hline
\multirow{3}{*}{B85.24} & 0.0160 & $0.4322(7)(^{+15}_{-2})$ & $0.4325(6)(^{+7}_{-0})$ \\
          & 0.0186 & $0.4442(7)(^{+15}_{-1})$ & $0.4441(6)(^{+8}_{-1})$ \\
          & 0.0210 & $0.4548(7)(^{+13}_{-0})$ & $0.4544(6)(^{+8}_{-1})$ \\
\hline
\multirow{3}{*}{D45.32} & 0.0130 & $0.2925(12)(^{+23}_{-0})$ & $0.2922(9)(^{+18}_{-0})$ \\
          & 0.0150 & $0.3028(11)(^{+5}_{-1})$ & $0.3010(9)(^{+14}_{-0})$ \\
          & 0.0180 & $0.3145(10)(^{+16}_{-0})$ & $0.3142(8)(^{+12}_{-0})$ \\
\hline
\multirow{3}{*}{D30.48} & 0.0115 & $0.2506(8)(^{+3}_{-5})$ & $0.2508(5)(^{+1}_{-5})$ \\
          & 0.0150 & $0.2677(8)(^{+3}_{-6})$ & $0.2679(6)(^{+1}_{-6})$ \\
          & 0.0180 & $0.2811(8)(^{+3}_{-6})$ & $0.2814(6)(^{+1}_{-7})$ \\
  \hline\hline
  \end{tabular*}
  \caption{Comparison of $E_{\pi K}$ for methods \E{1} and \E{2}}
  \label{tab:raw_data_energies_comp}
\end{table}
\begin{table}
  \centering
  \begin{tabular*}{.9\textwidth}{@{\extracolsep{\fill}}llll}
  \hline\hline
  Ensemble & $a\mu_{s}$ &  $a\delta E(\E{1})\times 10^{3}$ & $a\delta E(\E{2})\times 10^{3}$\\
\hline
\multirow{3}{*}{A30.32} & 0.0185 & $2.48(96)(^{+68}_{-20})$ & $2.44(81)(^{+51}_{-8})$ \\
          & 0.0225 & $2.41(97)(^{+81}_{-17})$ & $2.34(77)(^{+57}_{-8})$ \\
          & 0.0246 & $3.45(93)(^{+89}_{-49})$ & $3.14(74)(^{+19}_{-22})$ \\
\hline
\multirow{3}{*}{A40.24} & 0.0185 & $7.46(84)(^{+56}_{-0})$ & $6.93(56)(^{+43}_{-20})$ \\
          & 0.0225 & $6.70(85)(^{+97}_{-19})$ & $6.84(67)(^{+47}_{-1})$ \\
          & 0.0246 & $6.94(88)(^{+74}_{-0})$ & $6.69(61)(^{+50}_{-1})$ \\
\hline
\multirow{3}{*}{A40.32} & 0.0185 & $3.52(47)(^{+34}_{-83})$ & $3.15(32)(^{+0}_{-38})$ \\
          & 0.0225 & $3.42(45)(^{+85}_{-73})$ & $2.96(29)(^{+8}_{-27})$ \\
          & 0.0246 & $2.79(44)(^{+15}_{-42})$ & $2.99(31)(^{+13}_{-39})$ \\
\hline
\multirow{3}{*}{A60.24} & 0.0185 & $7.25(69)(^{+111}_{-0})$ & $7.02(39)(^{+80}_{-0})$ \\
          & 0.0225 & $8.05(46)(^{+24}_{-3})$ & $7.37(35)(^{+34}_{-36})$ \\
          & 0.0246 & $7.28(45)(^{+78}_{-43})$ & $7.22(30)(^{+46}_{-12})$ \\
\hline
\multirow{3}{*}{A80.24} & 0.0185 & $7.19(48)(^{+3}_{-22})$ & $6.45(21)(^{+30}_{-0})$ \\
          & 0.0225 & $6.47(47)(^{+48}_{-0})$ & $6.19(23)(^{+39}_{-0})$ \\
          & 0.0246 & $7.61(48)(^{+0}_{-57})$ & $6.40(21)(^{+34}_{-0})$ \\
\hline
\multirow{3}{*}{A100.24} & 0.0185 & $5.58(32)(^{+61}_{-30})$ & $5.70(17)(^{+27}_{-3})$ \\
          & 0.0225 & $5.68(32)(^{+48}_{-0})$ & $5.67(17)(^{+26}_{-0})$ \\
          & 0.0246 & $5.74(24)(^{+34}_{-30})$ & $5.69(14)(^{+19}_{-7})$ \\
\hline
\multirow{3}{*}{B35.32} & 0.0160 & $3.12(85)(^{+109}_{-0})$ & $3.38(45)(^{+62}_{-0})$ \\
          & 0.0186 & $3.94(54)(^{+23}_{-8})$ & $3.67(34)(^{+30}_{-13})$ \\
          & 0.0210 & $4.79(85)(^{+0}_{-33})$ & $3.72(56)(^{+53}_{-0})$ \\
\hline
\multirow{3}{*}{B55.32} & 0.0160 & $3.33(42)(^{+44}_{-12})$ & $3.71(30)(^{+28}_{-2})$ \\
          & 0.0186 & $3.02(45)(^{+64}_{-74})$ & $3.27(33)(^{+30}_{-8})$ \\
          & 0.0210 & $3.28(41)(^{+43}_{-12})$ & $3.65(30)(^{+27}_{-1})$ \\
\hline
\multirow{3}{*}{B85.24} & 0.0160 & $7.61(41)(^{+153}_{-16})$ & $7.88(23)(^{+66}_{-0})$ \\
          & 0.0186 & $8.05(36)(^{+150}_{-14})$ & $7.97(23)(^{+81}_{-5})$ \\
          & 0.0210 & $7.99(38)(^{+127}_{-3})$ & $7.64(25)(^{+81}_{-9})$ \\
\hline
\multirow{3}{*}{D45.32} & 0.0130 & $6.18(88)(^{+232}_{-0})$ & $5.87(53)(^{+177}_{-0})$ \\
          & 0.0150 & $7.62(86)(^{+54}_{-12})$ & $5.82(43)(^{+143}_{-0})$ \\
          & 0.0180 & $6.38(71)(^{+158}_{-0})$ & $6.05(45)(^{+120}_{-0})$ \\
\hline
\multirow{3}{*}{D30.48} & 0.0115 & $2.70(77)(^{+26}_{-53})$ & $2.90(50)(^{+10}_{-53})$ \\
          & 0.0150 & $2.79(79)(^{+26}_{-60})$ & $3.05(51)(^{+9}_{-63})$ \\
          & 0.0180 & $2.86(82)(^{+25}_{-65})$ & $3.17(54)(^{+9}_{-69})$ \\
  \hline\hline
  \end{tabular*}
  \caption{Comparison of $\delta E_{\pi K}$ for methods \E{1} and \E{2}}
  \label{tab:raw_data_deltae_comp}
\end{table}
\begin{table}
  \centering
  \begin{tabular*}{.9\textwidth}{@{\extracolsep{\fill}}llllll}
  \hline\hline
  Ensemble & $a\mu_s$ & $a_{0}/a(\E{1})$ & $a_{0}/a(\E{2})$ & $\ma^{3/2}(\E{1})$
  & $\ma^{3/2}(\E{2})$\\
\hline
\multirow{3}{*}{A30.32} & 0.0185 & $-0.96(34)(^{+7}_{-24})$ & $-0.94(29)(^{+3}_{-18})$ & $-0.077(27)(^{+6}_{-19})$ & $-0.076(23)(^{+2}_{-14})$ \\
          & 0.0225 & $-0.95(35)(^{+6}_{-29})$ & $-0.93(28)(^{+3}_{-20})$ & $-0.079(29)(^{+5}_{-24})$ & $-0.077(23)(^{+2}_{-17})$ \\
          & 0.0246 & $-1.33(32)(^{+17}_{-30})$ & $-1.23(26)(^{+8}_{-7})$ & $-0.112(27)(^{+14}_{-25})$ & $-0.103(22)(^{+6}_{-6})$ \\
\hline
\multirow{3}{*}{A40.24} & 0.0185 & $-1.27(12)(^{+0}_{-8})$ & $-1.19(8)(^{+3}_{-6})$ & $-0.114(11)(^{+0}_{-7})$ & $-0.107(8)(^{+3}_{-6})$ \\
          & 0.0225 & $-1.18(13)(^{+3}_{-15})$ & $-1.20(10)(^{+0}_{-7})$ & $-0.110(12)(^{+3}_{-14})$ & $-0.112(9)(^{+0}_{-7})$ \\
          & 0.0246 & $-1.23(14)(^{+0}_{-11})$ & $-1.20(10)(^{+0}_{-8})$ & $-0.116(13)(^{+0}_{-11})$ & $-0.112(9)(^{+0}_{-7})$ \\
\hline
\multirow{3}{*}{A40.32} & 0.0185 & $-1.42(17)(^{+31}_{-12})$ & $-1.29(12)(^{+14}_{-0})$ & $-0.126(15)(^{+27}_{-11})$ & $-0.114(10)(^{+12}_{-0})$ \\
          & 0.0225 & $-1.42(17)(^{+27}_{-31})$ & $-1.25(11)(^{+10}_{-3})$ & $-0.129(15)(^{+25}_{-28})$ & $-0.114(10)(^{+9}_{-3})$ \\
          & 0.0246 & $-1.20(17)(^{+17}_{-6})$ & $-1.28(12)(^{+15}_{-5})$ & $-0.111(16)(^{+15}_{-5})$ & $-0.118(11)(^{+14}_{-5})$ \\
\hline
\multirow{3}{*}{A60.24} & 0.0185 & $-1.37(11)(^{+0}_{-18})$ & $-1.33(6)(^{+0}_{-13})$ & $-0.139(11)(^{+0}_{-18})$ & $-0.135(6)(^{+0}_{-13})$ \\
          & 0.0225 & $-1.53(7)(^{+1}_{-4})$ & $-1.42(6)(^{+6}_{-6})$ & $-0.160(8)(^{+1}_{-4})$ & $-0.149(6)(^{+6}_{-6})$ \\
          & 0.0246 & $-1.43(7)(^{+7}_{-13})$ & $-1.41(5)(^{+2}_{-8})$ & $-0.151(8)(^{+8}_{-13})$ & $-0.150(5)(^{+2}_{-8})$ \\
\hline
\multirow{3}{*}{A80.24} & 0.0185 & $-1.48(8)(^{+4}_{-1})$ & $-1.35(4)(^{+0}_{-5})$ & $-0.165(9)(^{+4}_{-1})$ & $-0.150(4)(^{+0}_{-6})$ \\
          & 0.0225 & $-1.39(9)(^{+0}_{-9})$ & $-1.33(4)(^{+0}_{-7})$ & $-0.160(10)(^{+0}_{-10})$ & $-0.154(5)(^{+0}_{-8})$ \\
          & 0.0246 & $-1.61(8)(^{+10}_{-0})$ & $-1.39(4)(^{+0}_{-6})$ & $-0.188(10)(^{+12}_{-0})$ & $-0.162(5)(^{+0}_{-7})$ \\
\hline
\multirow{3}{*}{A100.24} & 0.0185 & $-1.27(6)(^{+6}_{-12})$ & $-1.29(3)(^{+1}_{-5})$ & $-0.153(8)(^{+7}_{-14})$ & $-0.156(4)(^{+1}_{-6})$ \\
          & 0.0225 & $-1.32(6)(^{+0}_{-9})$ & $-1.32(3)(^{+0}_{-5})$ & $-0.165(8)(^{+0}_{-12})$ & $-0.164(4)(^{+0}_{-7})$ \\
          & 0.0246 & $-1.35(5)(^{+6}_{-7})$ & $-1.34(3)(^{+1}_{-4})$ & $-0.170(6)(^{+8}_{-9})$ & $-0.169(4)(^{+2}_{-5})$ \\
\hline
\multirow{3}{*}{B35.32} & 0.0160 & $-1.14(28)(^{+0}_{-35})$ & $-1.22(15)(^{+0}_{-20})$ & $-0.088(22)(^{+0}_{-27})$ & $-0.095(11)(^{+0}_{-15})$ \\
          & 0.0186 & $-1.43(17)(^{+3}_{-7})$ & $-1.34(11)(^{+4}_{-10})$ & $-0.114(14)(^{+2}_{-6})$ & $-0.107(9)(^{+3}_{-8})$ \\
          & 0.0210 & $-1.73(26)(^{+10}_{-0})$ & $-1.38(19)(^{+0}_{-17})$ & $-0.140(21)(^{+8}_{-0})$ & $-0.112(15)(^{+0}_{-14})$ \\
\hline
\multirow{3}{*}{B55.32} & 0.0160 & $-1.38(15)(^{+4}_{-16})$ & $-1.52(11)(^{+1}_{-10})$ & $-0.125(14)(^{+4}_{-15})$ & $-0.137(10)(^{+1}_{-9})$ \\
          & 0.0186 & $-1.29(17)(^{+29}_{-24})$ & $-1.39(12)(^{+3}_{-11})$ & $-0.120(16)(^{+27}_{-22})$ & $-0.128(11)(^{+3}_{-10})$ \\
          & 0.0210 & $-1.42(16)(^{+5}_{-16})$ & $-1.55(11)(^{+0}_{-10})$ & $-0.133(15)(^{+4}_{-15})$ & $-0.146(10)(^{+0}_{-9})$ \\
\hline
\multirow{3}{*}{B85.24} & 0.0160 & $-1.47(7)(^{+3}_{-24})$ & $-1.52(4)(^{+0}_{-11})$ & $-0.155(7)(^{+3}_{-26})$ & $-0.160(4)(^{+0}_{-11})$ \\
          & 0.0186 & $-1.57(6)(^{+2}_{-24})$ & $-1.56(4)(^{+1}_{-13})$ & $-0.169(6)(^{+3}_{-26})$ & $-0.168(4)(^{+1}_{-14})$ \\
          & 0.0210 & $-1.59(6)(^{+1}_{-21})$ & $-1.53(4)(^{+2}_{-13})$ & $-0.174(7)(^{+1}_{-23})$ & $-0.167(5)(^{+2}_{-15})$ \\
\hline
\multirow{3}{*}{D45.32} & 0.0130 & $-1.89(23)(^{+0}_{-57})$ & $-1.81(14)(^{+0}_{-45})$ & $-0.132(16)(^{+0}_{-40})$ & $-0.126(10)(^{+0}_{-31})$ \\
          & 0.0150 & $-2.29(21)(^{+3}_{-13})$ & $-1.83(12)(^{+0}_{-37})$ & $-0.163(15)(^{+2}_{-9})$ & $-0.130(8)(^{+0}_{-26})$ \\
          & 0.0180 & $-2.02(19)(^{+0}_{-41})$ & $-1.94(12)(^{+0}_{-32})$ & $-0.149(14)(^{+0}_{-30})$ & $-0.142(9)(^{+0}_{-23})$ \\
\hline
\multirow{3}{*}{D30.48} & 0.0115 & $-2.43(60)(^{+43}_{-20})$ & $-2.58(38)(^{+42}_{-8})$ & $-0.144(36)(^{+25}_{-12})$ & $-0.152(23)(^{+25}_{-5})$ \\
          & 0.0150 & $-2.59(63)(^{+49}_{-20})$ & $-2.79(40)(^{+50}_{-7})$ & $-0.159(39)(^{+30}_{-12})$ & $-0.172(25)(^{+31}_{-4})$ \\
          & 0.0180 & $-2.71(66)(^{+54}_{-20})$ & $-2.95(42)(^{+56}_{-7})$ & $-0.171(42)(^{+34}_{-13})$ & $-0.187(27)(^{+36}_{-4})$ \\
  \hline\hline
  \end{tabular*}
  \caption{Comparison of $a_{0}$ and $\ma^{3/2}$ for methods \E{1} and \E{2}}
  \label{tab:raw_data_scatlen_comp}
\end{table}
\subsection{Interpolated Data}
\label{sec:chpt_A_input}
\begin{table}
  	\begin{tabular*}{.9\textwidth}{@{\extracolsep{\fill}}l
    S[round-mode=places,round-precision=2]
    S[round-mode=places,round-precision=3]
    S[round-mode=places,round-precision=3]
    S[round-mode=places,round-precision=3]
    S[round-mode=places,round-precision=3]} 
  	\hline\hline
	\centering
  Ensemble & {$\mu_{\pi K}/f_\pi$} & {$aM_K$} & {$aM_\eta$} &
  {$\ma^{3/2}$(\E{1})} & {$\ma^{3/2}$(\E{2})}  \\
\hline 
 A40.24 & 1.28 +- 0.02 & 0.241 +- 0.006 & 0.317 +- 0.006 & -0.12 +- 0.01 & -0.106 +- 0.008  \\ 
 A60.24 & 1.37 +- 0.01 & 0.251 +- 0.006 & 0.323 +- 0.006 & -0.15 +- 0.01 & -0.139 +- 0.009  \\
 A80.24 & 1.46 +- 0.01 & 0.260 +- 0.006 & 0.327 +- 0.005 & -0.16 +- 0.01 & -0.153 +- 0.006  \\
A100.24 & 1.52 +- 0.01 & 0.269 +- 0.006 & 0.332 +- 0.004 & -0.15 +- 0.01 & -0.158 +- 0.006  \\
 A30.32 & 1.22 +- 0.01 & 0.236 +- 0.006 & 0.314 +- 0.009 & -0.08 +- 0.03 & -0.08 +- 0.02    \\
 A40.32 & 1.28 +- 0.01 & 0.241 +- 0.006 & 0.314 +- 0.009 & -0.14 +- 0.02 & -0.11 +- 0.01    \\
 B85.24 & 1.49 +- 0.01 & 0.243 +- 0.005 & 0.296 +- 0.005 & -0.17 +- 0.02 & -0.178 +- 0.009  \\
 B35.32 & 1.27 +- 0.01 & 0.220 +- 0.006 & 0.284 +- 0.008 & -0.12 +- 0.02 & -0.11 +- 0.01    \\
 B55.32 & 1.40 +- 0.01 & 0.230 +- 0.005 & 0.283 +- 0.005 & -0.12 +- 0.03 & -0.14 +- 0.01    \\
 D45.32 & 1.45 +- 0.01 & 0.175 +- 0.004 & 0.200 +- 0.004 & -0.15 +- 0.02 & -0.15 +- 0.02    \\
 D30.48 & 1.29 +- 0.01 & 0.168 +- 0.004 & 0.196 +- 0.004 & -0.15 +- 0.03 & -0.15 +- 0.02   \\
\hline\hline
	\end{tabular*}
	\caption{Input Data for for the chiral analysis}
\end{table}
\begin{table}
	\centering
  	\begin{tabular*}{.95\textwidth}{@{\extracolsep{\fill}}l
	l
    S[fixed-exponent=-2,table-omit-exponent]
    S[fixed-exponent=-3,table-omit-exponent]
	S[round-mode=places,round-precision=3] 
    S[fixed-exponent=-2,table-omit-exponent]
	S[round-mode=places,round-precision=3]} 
  	\hline\hline
	ChPT &$E_{\pi K}$ & {$\ma^{3/2}\times 10^2$} & {$L_{\pi K}\times 10^3$} & {$\ma^{1/2}$} & {$M_\pi a_0^{3/2}\times 10^2$} & {$M_\pi a_0^{1/2}$}\\ 
	\hline
\multirow{2}{*}{$\Gamma$} & \E{1} &-0.047 +- 0.001  &  0.0037 +- 0.0002 &  0.128 +- 0.002  &  -0.060 +- 0.001  & 0.162 +- 0.002\\  
               			  		&	\E{2} &-0.0459 +- 0.0008  &0.0038 +- 0.0001 &0.129 +- 0.002  &-0.058 +- 0.001  & 0.164 +- 0.002\\ 
				 \hline                                                                                                                                                                     
\multirow{2}{*}{NLO} 	  	& \E{1} &  -0.047 +- 0.001  &  0.0036 +- 0.0002 &   0.127 +- 0.002&    -0.060 +- 0.001  &0.162 +- 0.002\\  
                		  		&	\E{2} &-0.0461 +- 0.0009  &  0.0038 +- 0.0002 &   0.129 +- 0.002&  -0.059 +- 0.001  &0.164 +- 0.002\\  
	\hline\hline
	\end{tabular*}
	\caption{Physical values of the scattering length and $L_{\pi K}$ after averaging over the fit ranges }
	\label{tab:phys_results_branches}
\end{table}
 \end{appendix}
\end{document}